\documentclass{aa}
\usepackage[utf8]{inputenc}
\usepackage{amssymb}
\usepackage{amsfonts} 
\usepackage{amsmath}
\usepackage{latexsym}

\usepackage{url}
\usepackage{xcolor}
\usepackage{pict2e}
\usepackage{hyperref}
\hypersetup{colorlinks,
citecolor=blue
}
\usepackage{graphicx,stackengine}
\usepackage[percent]{overpic}
\usepackage{graphicx}
\usepackage{subcaption}
\usepackage{hypcap}
\usepackage{wrapfig,booktabs}
\usepackage[export]{adjustbox}
\usepackage{needspace}
\usepackage{bibentry}
\usepackage{comment}
\usepackage{float}
\newcommand{\ie}{i.e.,}

\author{Anwesha Maharana}
\date{January 2022}

\newcommand{\degree}{\ensuremath{^\circ}}

\begin{document}

   %\title{Understanding the mispredicted space weather impact of the early September 2014 CMEs
   \title{Rotation and interaction of the September 8 and 10, 2014 CMEs tested with EUHFORIA}
   %{Unravelling the CME-CME interaction event of early September 2014 with data-driven EUHFORIA simulations}

   %\subtitle{I. Overviewing the $\kappa$-mechanism}

   \author{Anwesha Maharana
          \inst{1},\inst{2}
          \and
          Camilla Scolini\inst{3}
          \and
          Brigitte Schmieder\inst{1},\inst{4}
          \and
          Stefaan Poedts\inst{1},\inst{5}
          }

   \institute{Centre for mathematical Plasma Astrophysics (CmPA)/Dept.\ of Mathematics, KU Leuven, 3001 Leuven,
              Belgium\\
              \email{anwesha.maharana@kuleuven.be}
         \and
             Royal Observatory of Belgium, 1180 Uccle, Belgium\\
         \and
             Institute for the Study of Earth, Oceans and Space, University of New Hampshire, 03824 Durham, NH, USA\\
         \and
             LESIA, Observatoire de Paris, Universit\'e PSL , CNRS, Sorbonne Universit\'e, Universit\'e Paris-Diderot, 92190 Meudon, France\\
         \and
             Institute of Physics, University of Maria Curie-Sk{\l}odowska, 20-031 Lublin, Poland \\
          }

   \date{Received Month dd, yyyy; accepted Month dd, yyyy}

%Abstract has to be in curly braces
\abstract
{\textit{Context:} Solar coronal mass ejections (CMEs) can catch up and interact with preceding CMEs and solar wind structures to undergo rotation and deflection during their propagation. %while propagating through the heliosphere.
 \\ %chirality and orientation of the pre-eruptive features. The associated magnetic ejecta unexpectedly resulted in a positive $B_z$ at L1, implying a mismatch in the CME2 magnetic field orientation observed at the Sun and at L1. Instead, a geoeffective sheath was developed ahead of CME2 in the heliosphere that resulted in a minimum Dst $\sim-88$nT causing a moderate storm at Earth.\\ %Hence, the geoeffectiveness of the various sub-structures involved in this event was mispredicted. \\

\textit{Aim:} We aim to show how interactions undergone by a CME in the corona and heliosphere can play a significant role in altering its geoeffectiveness predicted at the time of its eruption. To do so, we consider a case study of two successive CMEs launched from the active region NOAA 12158 in early September 2014. %We study the evolution of two successive earthward CMEs that erupted from the active region AR 12158 on September 8, 2014, and September 10, 2014, respectively. 
%The first CME (CME1)
%second CME’s propagation 
%was not taken into consideration while forecasting and 
The second CME was predicted to be extensively geoeffective based on the remote-sensing observations of the source region. However, in situ measurements at 1~au recorded only a short-lasting weak negative $B_z$ component followed by a prolonged positive $B_z$ component.\\%The importance of constraining CME propagation in the heliosphere with appropriate observations. The case study of the early September 2014 CMEs, was challenging to understand as the space weather forecast went wrong. \\

\textit{Methods:} % to probe into the interesting phenomena of this event. 
The EUropean Heliosphere FORecasting Information Asset (EUHFORIA) is used to perform {{a self-consistent 3D MHD data-driven simulation of the two CMEs in the heliosphere. %The boundary conditions are obtained with test-independent EUHFORIA simulations of the background solar wind. The two CMEs are then inserted in the heliospheric domain using input parameters informed by observations near the Sun and fine-tuned to match in situ observations near 1 au. %CME1 is modelled with the LFF spheromak, and CME2 with the "Flux Rope in 3D" (FRi3D) model.
First, the ambient solar wind is modelled, followed by the time-dependent injection of CME1 with the LFF spheromak and CME2 with the "Flux Rope in 3D" (FRi3D) model. The initial conditions of the CMEs are determined by combining observational insights near the Sun, fine-tuned to match the in situ observations near 1~au, and additional numerical experiments of each individual CME.}}\\
%, and to evolve the flux rope CMEs (modelled with the LFF spheromak and the "Flux Rope in 3D" (FRi3D) models) in the heliosphere up to 2~au. \\

\textit{Results:} By introducing CME1 before CME2 in the EUHFORIA simulation, we modelled the negative $B_z$ component in the sheath region ahead of CME2 whose formation can be attributed to the interaction between CME1 and CME2. To reproduce the positive $B_z$ component in the magnetic ejecta of CME2, we had to initialise CME2 with an orientation determined at 0.1~au and consistent with the orientation interpreted at 1~au, instead of the orientation observed during its eruption. %This suggests a significant rotation of CME2 in the low corona.  
\\

\textit{Conclusions:} %(1) It is crucial to track CME rotation and deflection and reconstruct its orientation in the corona as far as possible before injecting it into CME evolution models like EUHFORIA for improving forecasting accuracy. 
EUHFORIA simulations suggest the possibility of a significant rotation of CME2 in the low corona in order to explain the in situ observations at 1~au. Coherent magnetic field rotations with enhanced strength (potentially geoeffective) can be formed in the sheath region as a result of interactions between two CMEs in the heliosphere even if the individual CMEs are not geoeffective.
}


   \maketitle
   
%%%---------------------%%%%%%%%%%%%%%%%%%%%%-------------------%%%

\section{Introduction}
\label{sec:intro}

\begin{figure}
    \centering
    \includegraphics[width=0.7\textwidth,trim={2cm 0.3cm 1cm 0cm},clip=]{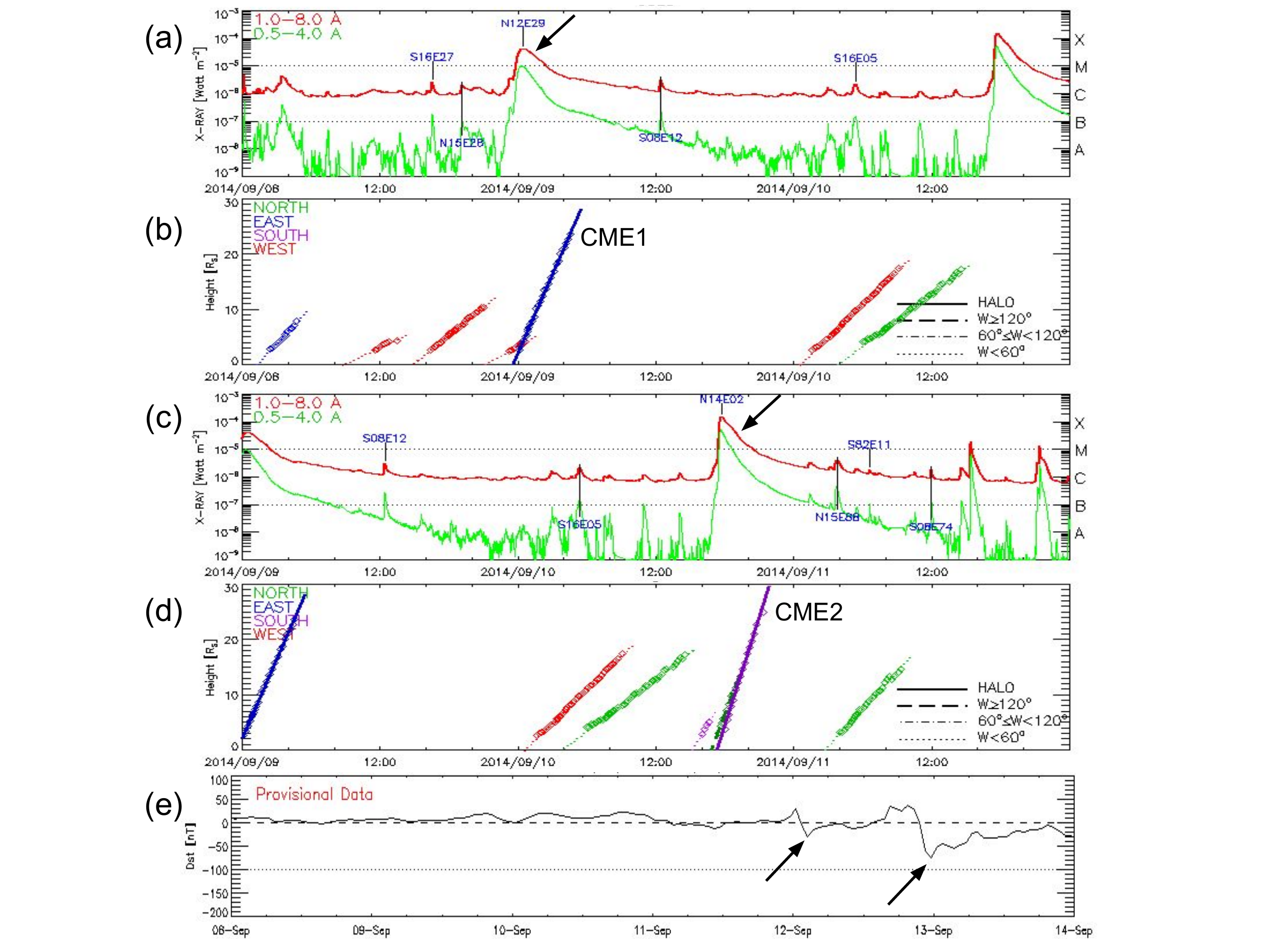}
    \caption{Overview of the eruption, early evolution in corona, and geomagnetic signatures of CME1 and CME2. Panels (a,b) and panels (c,d) span the time ranges of September 8-10, 2014 and September 9-11, 2014, respectively. (a) An M-class flare on late September 8, 2014 (indicated with black arrow), can be associated with the eruption of CME1; (b) Height-time plot of CME1 is shown by the blue height profile starting at $\sim$23.30~UT on September 8, 2014; (c) An X-class flare on September 10 (indicated with black arrow), can be associated with the eruption of CME2. (d) Height-time plot of CME2 is shown by the purple height profile starting $\sim$17.30~UT on September 9, 2014. The solar source coordinates of the flares are labelled in the GOES X-ray intensity plots in panels (a,c). The colour (line) codes in panels (b,d) define the CME propagation direction (apparent angular width). (e) Disturbance storm index (Dst), a measure of the geomagnetic activity at Earth shows a calm phase followed by a mild disturbance and then a moderate storm in the period 12-14 September 2014 (disturbances indicated with arrows). Source: CDAW catalogue - \url{https://cdaw.gsfc.nasa.gov/CME_list/daily_plots/sephtx/2014_09/}}
    \label{fig:cdaw_prop}
\end{figure}
%\textit{Outline: Talk about the uncertainties in space weather prediction. Phase 1: low corona observational limitation to determine magnetic field orientation in case of rotation; Phase 2: heliospheric deflection and lack of in situ observers to understand the global structure of the CME; Phase 3: inherent limitation of the operational forecasting models} \\
Coronal Mass Ejections (CMEs) can drive major geomagnetic storms \citep{gosling1991,Huttunen2005}. Hence, it is important to model their initiation and propagation in order to forecast their arrival at Earth or at any other planet or satellite. Uncertainties in space weather prediction are introduced by multiple factors, starting from the monitoring of the eruptions at the Sun, to the modelling of their propagation from the solar corona to the planets or satellites in the inner heliosphere \citep{riley2021,VERBEKE2022}. Magnetohydrodynamic (MHD) modelling is useful for tracking the propagation of CMEs, accounting for their interactions with solar wind structures and other CMEs, and computing their geoeffectiveness. Data-driven MHD modelling of CME evolution is more physical as it constrains the initial and boundary conditions using the early observations of the eruptions. However, if the orientation of the emerging CME is misinterpreted in the low corona, the initial conditions for the propagation models will yield inappropriate prediction results. In this work, we present and analyse such a case of space weather misprediction. %Fortunately, it was a false positive rather than a false negative which would have had worse consequences. There is a combination of the above drawbacks in this event that we  speculate and detail in this work. 
As part of the ISEST VarSITI campaign (\url{http://solar.gmu.edu/heliophysics/index.php/ISEST}), the CME event of September 10, 2014, was used to perform the exercise of real-time forecasting. The prediction was made considering the magnetic field signatures of the eruption on the solar surface and the direction of propagation estimated from the coronagraphic field of view. The CME was predicted to have a strong negative $B_z$ component and be a frontal impact at Earth \citep{Webb2017}. However, by the time the CME reached Earth, the associated magnetic ejecta \citep[ME;][]{Burlaga1988, Winslow2015} was characterised by a long-lasting positive $B_z$ component. A brief period of negative $B_z$ component was present in the sheath ahead of CME2 that drove a moderate storm (minimum Dst$\sim-88\;$nT) instead of the intense storm predicted, and hence, the geoeffectiveness of the different sub-structures associated with the CME was greatly mispredicted. Upon taking a closer look at this period, we noticed the presence of a preceding earthward CME that erupted late on September 8, 2014, and that was not recorded in any of the Interplanetary CME (ICME) catalogues at Earth. This preceding CME could precondition the propagation of the CME that erupted on September 10, 2014, and open the possibility of CME-CME interaction leading to the formation of the geoeffective sheath. %The possibility of CME-CME interaction affecting the formation of the sheath region ahead of a CME is also relevant. We explore different combinations of processes like CME rotation, interaction with other CMEs and solar wind structures during its propagation to Earth. \\
Specifically, we are seeking answers to two questions: 
(1) What is the orientation of the CME that erupted on September 10, 2014, at 0.1~au that must be injected in EUHFORIA in order to obtain the correct signature at 1~au?; and
(2) What is the role of the preceding CME in the formation of the negative $B_z$ component (or the magnetic field rotation) in the sheath region? 

The paper is organised as follows: Section~\ref{sec:event_overview} provides an observational overview of the event and builds the motivation of this study. In Sections~\ref{sec:obs_analysis_remote} and \ref{sec:obs_analysis_insitu}, we describe the event using various observational proxies, both remote and in situ, at the Sun ($1\;$R$_\odot$), close to 0.1~au and at 1~au. We perform MHD simulations of the event as described in Section~\ref{sec:euhforia}. Section~\ref{sec:results} {{presents the modelling results and our interpretation}} of this puzzling event, and Section~\ref{sec:conclusion} provides the {{summary and conclusions}}.  

%%%---------------------%%%%%%%%%%%%%%%%%%%%%-------------------%%%
\section{Observations} %{Overview of CMEs in the event}
\label{sec:event_overview}
%%%---------------------%%%%%%%%%%%%%%%%%%%%%-------------------%%%
\subsection{Overview of Sun-to-Earth signatures of the CMEs}
\begin{figure}
    \centering
    \includegraphics[width=0.5\textwidth]{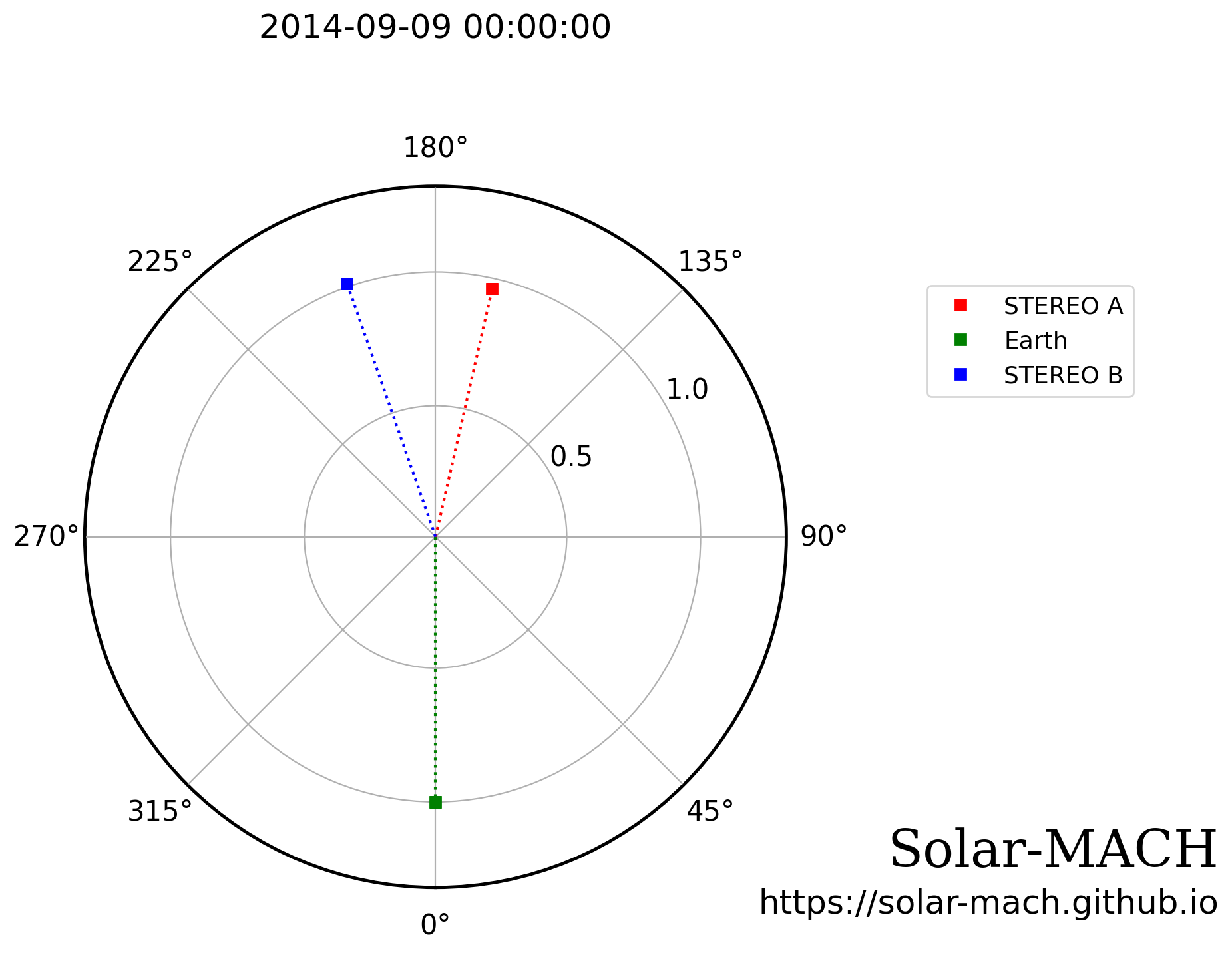}
    \caption{Position of STEREO-A, STEREO-B and Earth on September 9, 2014, at 00:00. The grid in black corresponds to the Stonyhurst coordinate systems. This polar plot is generated using the Solar-MACH tool \citep[\url{https://serpentine-h2020.eu/tools/};][]{Jan2022}.}
    \label{fig:sc_relative_pos}
\end{figure}

\begin{figure}
    \centering
    \includegraphics[width=0.24\textwidth]{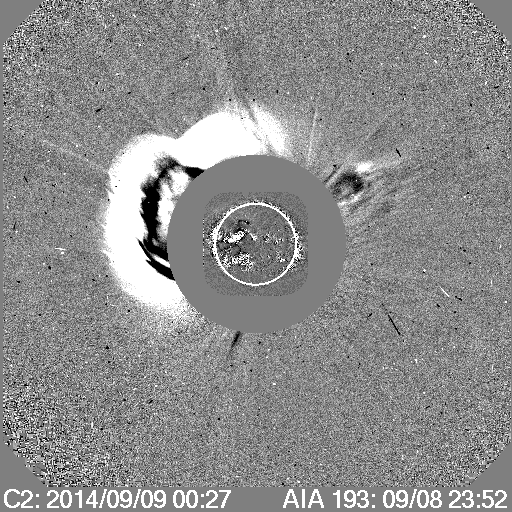}
    \includegraphics[width=0.24\textwidth]{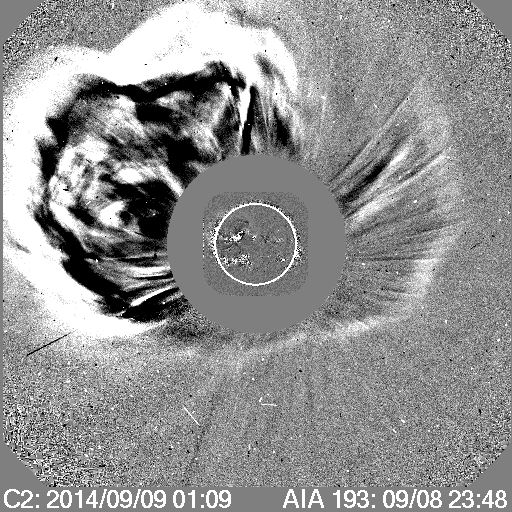} \\
    \includegraphics[width=0.24\textwidth]{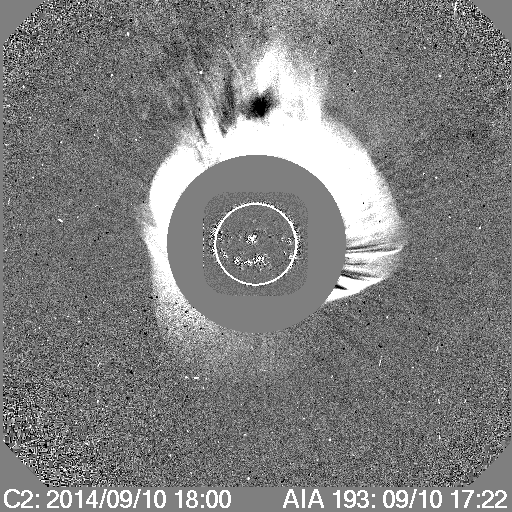}
    \includegraphics[width=0.24\textwidth]{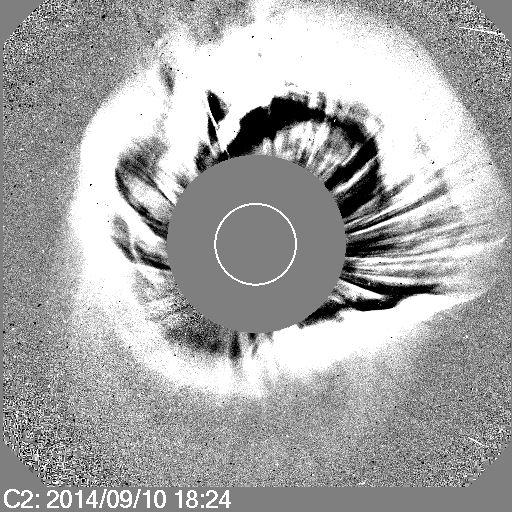}
    \caption{Running difference images showing the development of CME1 (top row) on early September 9, 2014, and CME2 (bottom row) on September 10, 2014, in the LASCO C2 field of view. Source: \url{https://cdaw.gsfc.nasa.gov/CME_list/}}
   \label{fig:cme_c2}
\end{figure}

\begin{table*}
\centering
\begin{tabular}{  p{5.0cm} | p{6.0cm} p{6.0cm}}
 \hline
 \hline
 %\multicolumn{3}{|c|}{\textbf{Input parameters}} \\
 %\hline
    &  CME1 & CME2 \\
 \hline
 \hline
 Active region position & N12E29 & N15E02\\
\hline
 Flare class   & M4.6 & X1.6\\
\hline
 Flare time (start - peak - end) & September 8, 23:12~UT - September 9, 00:28~UT - September 9, 01:30~UT & September 10, 17:21~UT - September 10, 17:45~UT - September 10, 17:45~UT \\
\hline 
 Apparent CME Speed [km~s$^{-1}$]   & $920$ &  $1267$\\
\hline
\hline
\end{tabular}
\caption{Observational details of the eruption of CME1 and CME2. Flare details are as reported on Solar Monitor (\url{https://www.solarmonitor.org/}). The position of the active region AR 12158, as per the NOAA catalogue, is in heliographic coordinates. CME speeds are taken from the CDAW catalogue as computed by an automatic linear fit.}
\label{tab:cme_obs}
\end{table*}

\begin{comment}
\begin{table}
\centering
\begin{tabular}{  p{3.5cm}| p{2.0cm} p{2.0cm}}
 \hline
 %\multicolumn{3}{|c|}{\textbf{Input parameters}} \\
 %\hline
    &  CME1 & CME2 \\
 \hline
 \hline
 Active region position & N12E29 & N15E02\\
\hline
 Flare class   & M4.6 & X1.6\\
 \hline
 Flare eruption (start - peak - end) & September 8, 2014 (23:12 UT) & September 10, 2014 (17:21 UT)\\
 \hline
 Flare eruption (peak) & September 9, 2014 (00:28 UT) & September 10, 2014 (17:45 UT)\\
 \hline
 Flare eruption (stop) & September 9, 2014 (01:30 UT) & September 10, 2014 (17:45 UT)\\
 \hline
 Apparent CME Speed [km~s$^{-1}$]   & $920$ &  $1267$\\

\hline
\end{tabular}
\caption{Observational details of both CMEs. Flare details are as reported in the Solar monitor (\url{https://www.solarmonitor.org/full_disk.php?date=20140910&type=shmi_maglc&region=&indexnum=1}). The position of the active region (AR 12158) as per NOAA catalogue is in heliographic coordinates. CME speeds are taken from CDAW catalogue.}
\label{tab:cme_obs}
\end{table}
\end{comment}
%https://www.frontiersin.org/articles/10.3389/fspas.2021.631582/full
%CME1: Flare erupted on September 8, 2014 at N12E29 at 23:

In this section, we identify the observational signatures of the two successive CMEs that occurred between September 8-10, 2014. The first CME (hereafter, CME1) was associated with an M4.6 flare occurring in the Active Region NOAA 12158 (hereafter, AR 12158), positioned at N12E29, on September 8, 2014, starting around 23:12~UT. The flare peaked at 00:28~UT on September 9, 2014. The origin of the second CME (hereafter, CME2) has been extensively studied \citep{Cheng2015,Dudik2016,Zhao2016}. CME2 was associated with an X1.6 flare which started on September 10, 2014, at 17:21~UT from AR 12158, positioned at N15E02, and peaked at 17:45~UT. Figure~\ref{fig:cdaw_prop}(a) and ~\ref{fig:cdaw_prop}(c) indicate the flares and provide the X-ray intensities associated with the eruption of CME1 and CME2, respectively. During the early propagation of CME1 and CME2, they were detected in the field of view (FOV) of the C2 and C3 instruments of Large Angle and Spectrometric COronagraph \citep[LASCO,][]{Brueckner1995} on board the Solar and Heliospheric Observatory (SOHO), and the COR-2 instrument on board the Sun-Earth Connection Coronal and Heliospheric Investigation (SECCHI) package of the twin-spacecraft Solar Terrestrial Relations Observatory \citep[STEREO,][]{Kaiser2008}. Only STEREO-B recorded the observations while a data gap was found in STEREO-A during this period. Figure~\ref{fig:sc_relative_pos} shows the relative positioning of the observing spacecraft in the heliosphere. CME1 was visible in the C2 FOV at 00:06~UT with an apparent speed of $920\;$km~s$^{-1}$ and in COR-2B FOV at 00:24~UT. CME2 was first observed by C2 at 17:48~UT, and it developed as a halo CME at 18:24~UT. It was later visible in the C3 FOV starting from around 18:45~UT with an apparent speed of $1267\;$km~s$^{-1}$. CME2 appeared for the first time by the COR-2B at 18:24~UT. The height-time profiles of CME1 and CME2 up to $\sim 30\;$R$_\odot$, created by automatic tracking of the CME leading edge and approximating a linear fit by the CDAW catalogue are shown in Fig.~\ref{fig:cdaw_prop}(b) and \ref{fig:cdaw_prop}(d), respectively. %CME2 appeared for the first time in the COR-1B and COR-2B FOV at 17:45~UT and 18:10~UT respectively. %At the time, STEREO-B was located at heliographic (HEEQ) latitude and longitude of $-6.9\degree$ and $-161\degree$ respectively. 
%\url{https://secchi.nrl.navy.mil/cactus/index.php?p=SECCHI-B/2014/09/out/latestCMEs.html}
%CME2: 2014/09/10 17:24| 04 | 330| 338| 0892| 0176| 0416| 1136|  IV
%CME1: 2014/09/09 00:24| 05 | 313| 180| 0735| 0099| 0480| 0961|  II
The above details are also listed in Table~\ref{tab:cme_obs}. The association of the CMEs with the corresponding flares is also reported by \citet{Vemareddy2016}. Figure~\ref{fig:cme_c2} shows the evolution of both the CMEs in the C2 FOV. %CME1 must arrive at 1~au in $45$ hours, \ie on $\sim$September 10, 20:00 and CME2 on $\sim$September 12, 01:00.
CME2, tagged as a textbook event \citep{Webb2017}, reached Earth on September 12, 2014. The arrival of CME2 at L1 is recorded in the WIND ICME catalogue (\url{https://wind.nasa.gov/ICME_catalog/ICME_catalog_viewer.php}) at 15:17~UT and in the Richardson and Cane catalogue \citep[\url{https://izw1.caltech.edu/ACE/ASC/DATA/level3/icmetable2.htm};][]{Cane2003,Richardson2010} at 15:53~UT, respectively. The interplanetary counterpart of CME1 is not listed in any of the ICME catalogues. No other wide and Earthward CMEs were observed between the time period starting from the eruption of CME1 until two days after the eruption of CME2, which could have arrived at L1 interrupting or affecting the two CMEs of our concern. Extrapolating the CME arrival times at Earth using the projected speeds from the CDAW catalogue (\url{https://cdaw.gsfc.nasa.gov/CME_list/}), we obtain a time interval between their estimated arrival times of about 29 hours. This is a rough estimate assuming no effects from the drag and the interaction of the CMEs in the heliosphere affect their kinematics during propagation. The time difference between the arrival times observed in situ at L1 is $\sim 32$ hours, hence corroborating the calculated time difference and the association between the coronal and interplanetary signatures of the eruptions. Signatures of both the CMEs were identified in the Disturbance storm index (Dst) at Earth as shown in Fig.~\ref{fig:cdaw_prop}(e). A low drop in Dst ($\sim-40\;$nT) followed by a moderately negative storm (Dst$\sim -88\;$nT) was observed. This prompted the preliminary association of the CMEs in the low corona and the CMEs at 1~au. \\%However, the CME was initially predicted to have a strong southward magnetic field component (negative $B_z$) which would have resulted in a strong geomagnetic storm, assuming a direct hit at Earth. \\

\subsection{In situ signatures at Earth}
\begin{figure*}[ht!]
    \centering
    \includegraphics[width=0.8\textwidth]{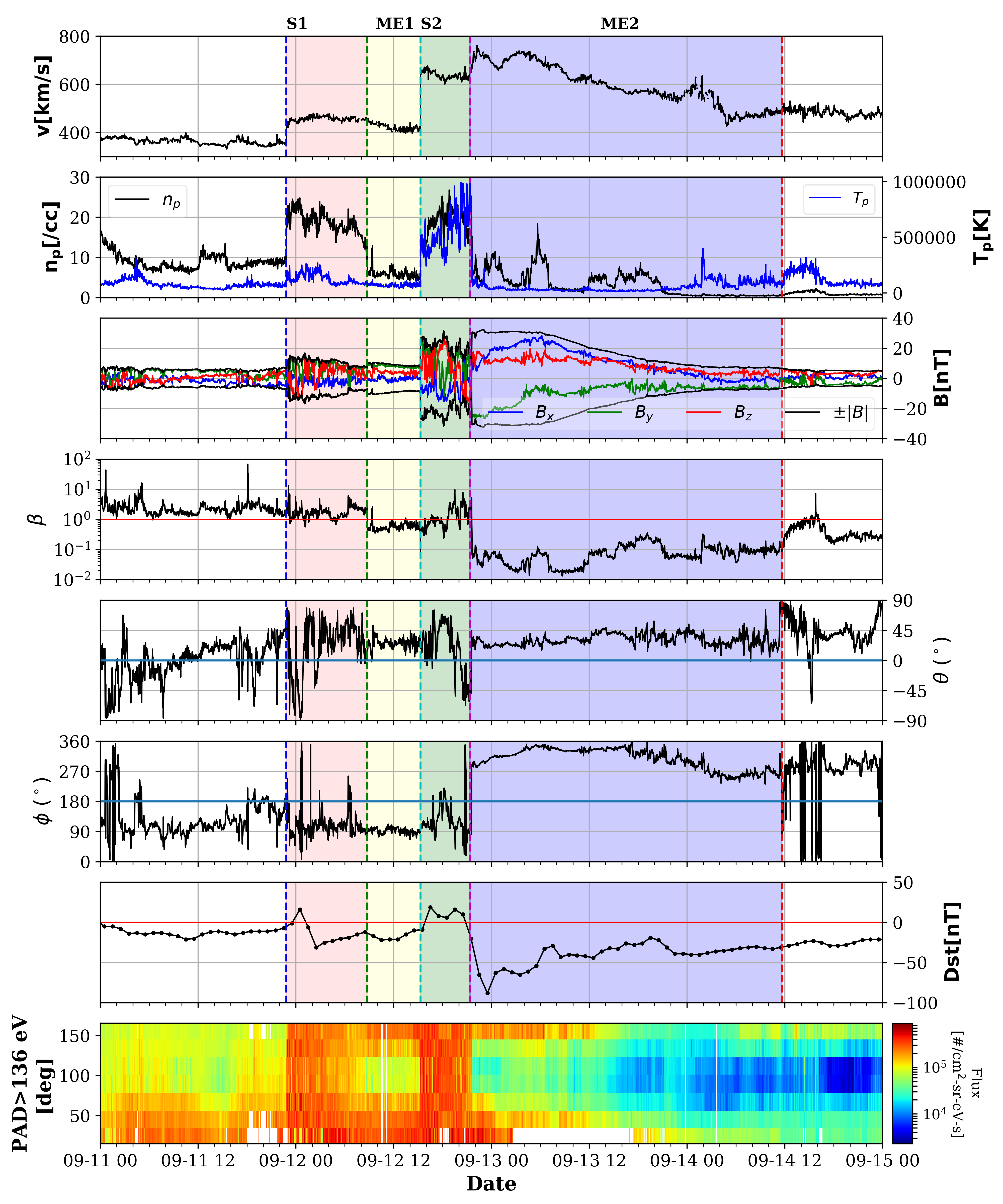} %insitu_plot.png}
    \caption{In situ measurements by the Wind spacecraft during the period of September 11, 2014, and September 15, 2014. The figure shows (top to bottom) speed ($v$), proton temperature ($T_p$) and number density ($n_p$) in the second panel, magnetic field components ($B_x$, $B_y$, $B_z$) in the GSE coordinate system capped with the total magnetic field ($\pm|B|$) in the third panel, total magnetic field ($|B|$), the plasma beta ($\beta$) in the fourth panel, the $\theta$ and $\phi$ components of the magnetic field in GSE angular coordinates in the fifth and sixth panel, respectively, the Dst index in the seventh panel, and in the last panel the suprathermal pitch angle distribution (with energies $>136\;$eV). Vertical dashed lines indicate the shock arrival of CME1 (S1, blue) and the start of the magnetic ejecta passage of CME1 (ME1, green). The shock arrival of CME2 (S2, cyan), the start of the magnetic ejecta passage of CME2 (magenta) and the end of the magnetic ejecta passage (red) are as identified in the Wind ICME catalogue. The shaded red and yellow regions represent the sheath ahead of CME1 and the magnetic ejecta, ME1, respectively. The shaded green and blue regions depict the sheath ahead of the CME2 and the magnetic ejecta, ME2. The magnetic ejecta of CME2 has been identified in the Richardson and Cane catalogue as well.}
    \label{fig:wind_insitu}
\end{figure*}

The in situ signatures of the CMEs are plotted in Fig.~\ref{fig:wind_insitu}. The shock (S1) driven by CME1 is observed on September 11 at 22:50~UT based on the IPShock catalogue \citep[\url{http://ipshocks.fi/};][]{Kilpua2015}. %IPShock catalogue\footnotemark{}\footnotetext{IPShock catalogue - \url{http://ipshocks.fi/}}\citep{Kilpua2015}. 
It is characterised by the enhancement in speed and number density. CME1 is directed north of the ecliptic plane as seen from the coronagraph images from LASCO instruments\footnotemark{}\footnotetext{CDAW animation - \url{https://cdaw.gsfc.nasa.gov/movie/make_javamovie.php?stime=20140908_2258&etime=20140909_0313&img1=lasc2rdf&title=20140909.000626.p059g;V=920km/s}}. The Space Weather Database Of Notifications, Knowledge, Information (DONKI, \url{https://kauai.ccmc.gsfc.nasa.gov/DONKI/search/}) catalogue has also recorded a north-eastward launch direction of the CME. These observations suggest that the WIND spacecraft would have encountered the southwestern flank of CME1. The long sheath region (characterised by density enhancement and fluctuating magnetic field signatures in the red-shaded region in Fig.~\ref{fig:wind_insitu}) after the shock of CME1 can also be inferred as the signature of the CME1 flank. Following these clear sheath signatures, we observe a simultaneous decrease in density and temperature, plasma beta ($\beta$) being less than 1, and a slight apparent increase in the magnetic field after the turbulent phase. However, the lack of a clear rotation in the magnetic field vector suggests the passage of the CME1 flank. A period of bi-directional electrons can also be observed between S1 and S2 (corresponding to the yellow-shaded region), suggesting their propagation inside the magnetic ejecta (hereafter, ME1) associated with CME1. Although some other typical characteristics of MEs, such as a significantly enhanced magnetic field, clear rotations in the magnetic field components, and oxygen enhancements, are missing, the low density, temperature, $\beta$, and magnetic field fluctuations in combination with bi-directional electrons suggest the passage of ME1 associated with CME1 starting on September 12 at 8:45~UT as indicated in the yellow shaded region \citep{zurbuchen2006}. The second shock (S2) is recorded on September 12 at 15:17~UT. S2 is followed by a distinct turbulent sheath (green shaded region) and a clear magnetic ejecta (hereafter, ME2, shaded blue region). The start and end times of ME2, as recorded in the WIND catalogue, are September 12 at 21:22~UT and September 14 at 11:38~UT, respectively. The Richardson and Cane catalogue reports the ME2 boundaries as September 12 at 22:00~UT and September 14 at 2:00~UT. Upon closer visual inspection of the data, we find that the ME2 boundaries from the WIND catalogue include some part of the sheath before ME2, and the boundaries listed in the Richardson and Cane catalogue detect the end of ME2 $\sim 10\;$hours earlier than the WIND catalogue. Although the WIND catalogue visibly provides better boundaries for ME2, we correct the start time to September 12 at 21:36~UT based on the visual inspection to remove parts of the sheath. As seen in Fig.~\ref{fig:wind_insitu}(panel 1), a decreasing speed profile within ME2 points to an expanding structure passing through the spacecraft. Inside ME2, the plasma density and temperature are also reduced (Fig.~\ref{fig:wind_insitu}, panel 2). Rotations in the magnetic field direction, an enhancement in the magnetic field strength, and a reduced $\beta$ can also be observed in the shaded blue region in Fig.~\ref{fig:wind_insitu}(panels 3, and 4). The $\theta$-profile (Fig.~\ref{fig:wind_insitu}, panel 5) does not show a smooth rotation in the north-south magnetic field direction (i.e., $|\Delta\theta|\neq 0$), but rather a constant and long-duration positive profile. This observation is compatible with the passage of a CME flank. There is a jump in $\phi$ from $90\degree$ to $> 180\degree$ (Fig.~\ref{fig:wind_insitu}, panel 6) which implies a westward axis of CME2. \citet{Marubashi2017} performed a toroidal CME model fitting to the in situ measurements of ME2 and reported a small rotation angle of the observed magnetic field vector. Their most important takeaway was that, although the magnetic field orientation of the CME2 is southward, the observed northward magnetic field inside ME2 could be due to the impact of the southern edge of the CME as the CME propagation was mainly north of the ecliptic. This highlights how crucial it is to predict what part of the gigantic CME would impact Earth in order to forecast the geomagnetic effects. A period of bi-directional suprathermal electrons corresponding to the ME2 boundaries can be observed (Fig.~\ref{fig:wind_insitu}, panel 8). Other magnetic ejecta signatures, such as an enhanced oxygen charge ratio (O$^{+7}$/O$^{+6}$) and average iron charge ratio $<Q_F>$ have also been reported in association with ME2 in \citet{Kilpua2021}. The geomagnetic storms at L1 during the period of September 12-14, 2014 are characterised by the negative Dst index (Fig.~\ref{fig:wind_insitu}, panel 7). The negative $B_z$ component in the sheaths ahead of CME1 and CME2 can be associated with the weak storm on September 12 and the moderate storm on September 13, respectively. \newline \newline% The weak storm on September 12 (moderate storm on September 13) can be associated with the negative $B_z$ component in the sheath ahead of CME1 (CME2). %and they can be associated with the negative $B_z$ component in the sheath ahead of CME2.

\citet{Kilpua2021} modelled CME2 with a time-dependent magnetofrictional model initialised with a flux rope about the PIL as suggested by \citet{Vemareddy2016} and with a chirality consistent with the inverse S-shaped EUV sigmoid. %The core dimming regions (See Fig.~8(c) in \citet{Kilpua2021}) in their analysis confirm the axial magnetic field orientation of the erupting flux rope to be Eastward. The sigmoid orientation along with the axial field direction points to a southward followed by a northward poloidal component ($B_z$) if the CME impact is head-on at 1~au. However, even by taking measurements for different lineouts to represent CME crossing the Earth with different parts, the $B_z$ predicted by extrapolating the TMFM simulations up to 1~au, could not match the in situ observations. 
The modelling results did not match the in situ magnetic field observations. Even if they inferred that CME2 was a flank hit at Earth, the extrapolated $B_z$ component was still mainly negative contrary to the in situ observations at Earth. Some studies have shown that initializing CME2 in MHD heliospheric propagation models using the PIL orientation as in \citet{Vemareddy2016}, did not match the magnetic field configuration at 1~au \citep{An2019}. The results from the previous studies raise the question of whether the CME rotated further anti-clockwise, either in the low corona or during heliospheric propagation in a way that could have led to the passage of a westward axial magnetic field with a dominant positive $B_z$ component through Earth.
\section{Reconstruction of the CMEs from remote-sensing observations in the corona}
\label{sec:obs_analysis_remote}
In this section, we derive the magnetic, geometric, and kinematic CME parameters from remote-sensing observations which we will later use to initialise the CMEs in the heliospheric simulations. In this work, the CMEs are modelled as magnetic flux ropes \citep[defined as bundles of twisted magnetic field lines with electric fields flowing inside;][]{antiochos1999,Torok2005}. First, the chirality and the orientation of the erupting flux rope are constrained from the CME source region proxies in Section~\ref{subsec:mag_sign_sun}. Second, the magnetic flux is derived using statistical relations based on the X-ray flare intensity in Section~\ref{sec:mag_reconstruct_sun}. Finally, a 3D geometrical reconstruction is performed using remote sensing coronagraph observations from LASCO and STEREO-B in the corona below 0.1~au in Section~\ref{subsec:geo_reconstruct}. %Large Angle and Spectrometric Coronagraph Experiment (LASCO) on Solar and Heliospheric Observatory (SOHO) and Sun Earth Connection Coronal and Heliospheric Investigation (SECCHI) onboard Solar Terrestrial Relations Observatory (STEREO).
%%%---------------------%%%%%%%%%%%%%%%%%%%%%-------------------%%%
\subsection{Source region observations}
\label{subsec:mag_sign_sun}
%%%---------------------%%%%%%%%%%%%%%%%%%%%%-------------------%%%

\begin{comment}
    \begin{overpic}[width=0.32\textwidth,trim={1.2cm 3.0cm 2.5cm 2.2cm},clip=]{Figures/CME1/max1000_aia131_2014-09-08T23:37:32.png}
    \put (10,80) {\color{white}{\textbf{(a)}}}
    \end{overpic}
    
    \begin{overpic}[width=0.32\textwidth,trim={1.2cm 3.0cm 2.5cm 2.2cm},clip=]{Figures/CME1/max10000_aia171_2014-09-09T00:15:11.png}   
    \put (10,80) {\color{white}{\textbf{(b)}}}
    \end{overpic}
    
    \begin{overpic}[width=0.32\textwidth,trim={1.2cm 3.0cm 2.5cm 2.2cm},clip=]{Figures/CME1/max10000_aia304_2014-09-09T00:15:07.png}%max500_aia1600_2014-09-09T00:15:04.png}
    \put (10,80) {\color{white}{\textbf{(c)}}}
    \end{overpic}
    
    \begin{overpic}[width=0.32\textwidth,trim={1.2cm 3.0cm 2.5cm 2.2cm},clip=]{Figures/t45_pil_hmi_2014-09-09T00:15:36.png}
    \put (10,80) {\color{white}{\textbf{(d)}}}
    \end{overpic}
\end{comment}

\begin{figure}
    \centering
    %{\includegraphics[width=0.5\textwidth,trim={7cm 3cm 7cm 4.0cm},clip=]{Figures/CME1_solar_images1.pdf}}
    {\includegraphics[width=0.5\textwidth,trim={3.3cm 0cm 4cm 0cm},clip=]{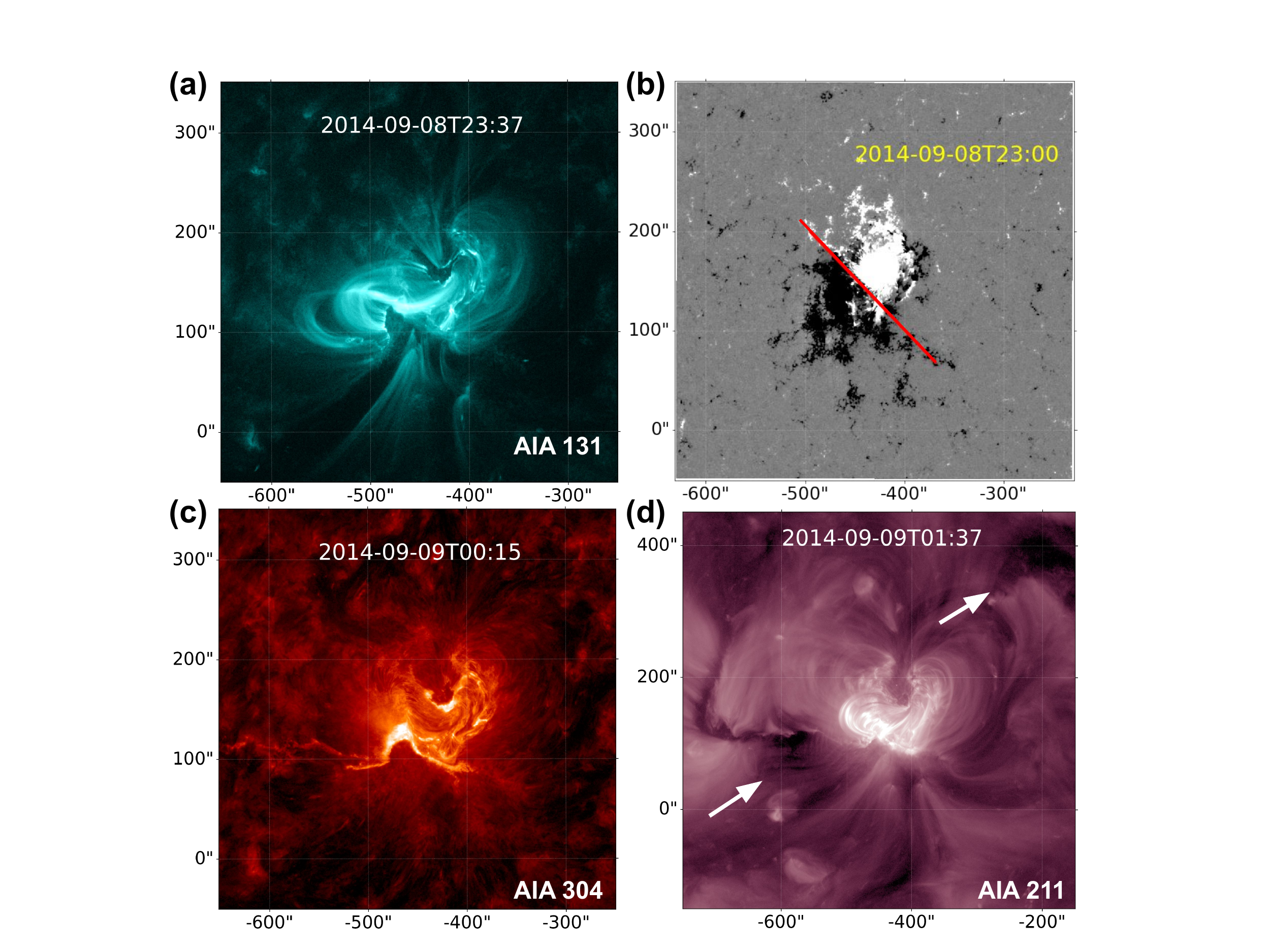}} %CME1_Sept2014_updated-1.pdf
    \caption{AR 12158 associated with CME1 eruption in different wavelengths - (a) AIA~$131\;$\AA \ image highlights the evolved sigmoid and the hooks corresponding to flux rope footpoints during the early phase of the flare; (b) HMI magnetogram saturated at $\pm$200~G overlaid with the approximated PIL orientation (i.e., the part of the extended PIL that most likely erupted as CME1), before the start of the flare; (c) AIA~$304\;$\AA \ image shows the inverse J-shaped flare ribbons after the eruption, which suggest the eruption of a left-handed flux rope, and (d) AIA~$211\;$\AA \ image shows the post flare coronal dimmings (marked with white arrows). The $X-$ and $Y-$axes correspond to the helio-projective longitude and latitude, respectively.}
    \label{fig:cme1_source}
\end{figure}

\begin{figure*}
    \centering
    {\includegraphics[width=0.9\textwidth,trim={0cm 13cm 0 0cm},clip=]{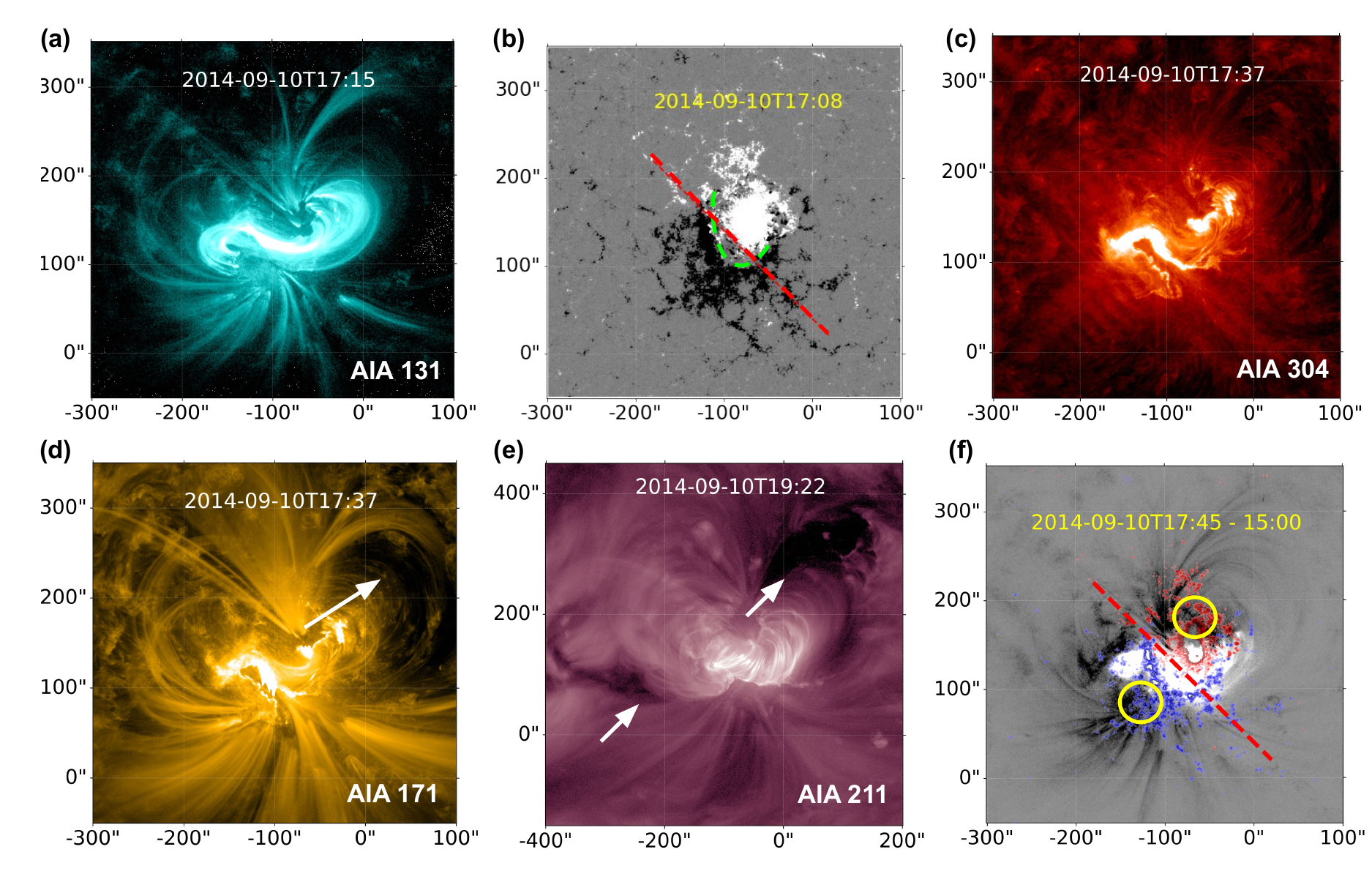}}
    \caption{AR 12158 associated with CME2 eruption in different wavelengths - (a) AIA~$131\;$\AA \ image before CME2 eruption; (b) HMI magnetogram saturated at $\pm$200~G overlaid with the approximated PIL drawn with a red dashed line; (c) AIA~$304\;$\AA \ image highlights the inverse J-shaped flare ribbons indicating left-handedness of the flux rope; (d) AIA~$171\;$\AA \ image showing the coronal loops and the eruption direction of CME2 in north-westward direction as per \citet{Dudik2016}; (e) AIA~$211\;$\AA \ image showing coronal dimmings marked with white arrows; (f) AIA~$131\;$\AA \ base-difference image overlaid with HMI magnetogram contours saturated at $\pm1000\;$G coloured blue (red) for negative (positive) polarity. The yellow circles demarcate the dimmings located at the footpoints of the FR. The red dashed line is the approximated PIL. The $X-$ and $Y-$axes correspond to the helio-projective longitude and latitude, respectively.}
    \label{fig:AR_wv_hmi}
\end{figure*}
In this section, the pre-, during and post-eruptive magnetic field signatures are analysed with remote sensing observations from the Atmospheric Imaging Assembly (AIA) and Helioseismic and Magnetic Imager (HMI) on board the Solar Disk Observatory \citep[SDO;][]{Lemen2012}. AR 12158 appeared rotated onto the disc around September 3, 2014, and erupted twice, once on September 8 and again on September 10. Observations of the active region in different wavelengths during the eruption of CME1 and CME2 are shown in Fig.~\ref{fig:cme1_source} and \ref{fig:AR_wv_hmi}. We place our focus mainly on CME2 in order to understand its orientation during eruption and to probe the reasons for possible rotations in the low corona which could have led to the mismatch of its orientation at 1~au. To estimate the magnetic field orientation of the front part of the flux rope, we investigate the magnetic chirality and polarity of the active region and the orientation of the polarity inversion line (PIL). In addition, we check the associated dimming regions to identify the footpoints of the erupting flux rope to further support the orientation inferred from the PIL. 
%On September 10, the positive polarity central spot of the active region was surrounded by an extended diffused negative region as shown in Fig.~\ref{fig:AR_wv_hmi}(b). The roots of the observed inverse S-shaped sigmoid were located on the rotating sunspot in the active region \citep{Vemareddy2016}. \citet{Dudik2016} point out the peculiarity of the active region in terms of having a positive polarity leading sunspot. During cycle 24, the leading sunspots in the northern hemisphere are expected to have negative polarity. %This section is mainly concerned with the eruption of CME2.

\textit{Flux rope chirality (the sign of magnetic helicity)}: 
 %A detailed analysis of various proxies proposed by \citet{Palmerio2017} to determine the chirality of an active  region  has been achieved. (i) In HMI images the elongations of the main magnetic polarities formed during the emergence of a flux tube through the photosphere, called magnetic tongues, can characterise the chirality based on the positive or negative twist.
Using the EUV/soft X-ray sigmoid as a proxy for an emerging flux rope embedded in an arcade \citep{Titov1999}, a reverse S-shaped sigmoid is identified in AIA~131~\AA \ before the eruption of CME1 and CME2 as shown in Fig.~\ref{fig:cme1_source}(a) and \ref{fig:AR_wv_hmi}(a), respectively. This suggests the erupting flux ropes associated with both CME1 and CME2 likely have a left-handed chirality. The leading positive magnetic polarity extends over the southern part of the trailing polarity and gives the active region a negative twist \ie \ a proxy for left-handed chirality \citep{Luoni2011} as shown in the HMI magnetogram images in Fig.~\ref{fig:cme1_source}(b) and \ref{fig:AR_wv_hmi}(b). Additionally, the sigmoids were well identified with two inverse J-shaped flare ribbons in AIA~$304\;$\AA \ characterising their left-handedness in Fig.~\ref{fig:cme1_source}(c) and Fig.~\ref{fig:AR_wv_hmi}(c) for CME1 and CME2 respectively \citep{Palmerio2017}.

\textit{Flux rope orientation}: The orientation (tilt) of the erupting flux rope is inferred from the orientation of the polarity inversion line (PIL), which is usually parallel to the axial magnetic field of the flux rope. The tilt angle, measured from the solar west is assigned a positive (negative) value if the acute angle is calculated counter-clockwise (clockwise) from the ecliptic on the solar west. The directionality of the axis is determined from the chirality of the active region. % \citep{Wang2013,Marubashi2015}. 
The flare associated with CME1 was localised to the eastern part of the extended PIL which could be approximated with a straight line as shown in Fig.~\ref{fig:cme1_source}(b).
%Although approximating the PIL with a straight line works for CME1 (Fig.~\ref{fig:cme1_source}(b)), it is not straightforward in the case of CME2 because of the curved geometry of the PIL (green dashed line in Fig.~\ref{fig:AR_wv_hmi}(b)). 
However, determining a univocal orientation for the PIL in the case of CME2 was not straightforward because the eruption extended along the curved geometry of the PIL (green dashed line in Fig.~\ref{fig:AR_wv_hmi}(b)). In the case of both CME1 and CME2, we consider the main axial field direction as northeastward making an angle $\sim -45\degree$ with the ecliptic.  %The inclination of the pair of footpoints seems to decrease anti-clockwise (from 17.15~UT to 18.00~UT). 
The magnetic field topology of AR 12158 reconstructed with a nonlinear force-free field (NLFFF) model also corroborates the presence of a highly twisted pre-eruptive flux rope surrounded by inverse J-shaped magnetic field lines \citep{Zhao2016}. \citet{Dudik2016} found evidence of the occurrence of slipping reconnection in the flaring region where flare loops slip towards both ends of the ribbons. When the eruption occurs, the filaments are observed getting disturbed in AIA~$171\;$\AA \ in a northwestward direction, which is also identified by \citet{Dudik2016} as indicated by the white arrow in Fig.~\ref{fig:AR_wv_hmi}(d). %A large dimming region is observed northwest of the active region pointing to the evacuation of matter due to the deflected rising flux rope. 
The location of the footpoints of the flux rope is also identified with the coronal dimming signatures in AIA~$211\;$\AA as shown in the white arrows in Fig.~\ref{fig:AR_wv_hmi}(e). In Fig.~\ref{fig:AR_wv_hmi}(f), the base difference image in AIA $131\;$\AA \ overlaid with HMI magnetogram (saturated at $\pm 1000\;$G; blue for positive and red for negative polarity) after the CME2 eruption. The development of the dark dimmings was observed in the southeast and northwest parts of the active region lying in the negative and positive magnetic polarity regions as marked by yellow circles. This suggests the eruption of the flux rope almost parallel to the linear PIL marked by the red dashed line. The orientation of the main PIL of AR 12158 associated with the CME1 and CME2 eruptions is consistent with the descriptions provided by \citet{Vemareddy2016, Dudik2016, Zhao2016}, \ie the tilt is $\sim -45\degree$ using straight-line assumption.

The conclusions drawn from the observations of the flare-CME eruption phase are as follows: (a) The axial flux rope fields of the CMEs are directed eastward; (b) The CMEs were characterised by a left-handed helicity, which combined with the eastward axial fields implies north to south poloidal field lines at the flux rope apex, hence characterising the magnetic topology of the flux ropes as SEN \citep{Bothmer1998}.

%%%---------------------%%%%%%%%%%%%%%%%%%%%%-------------------%%%
%In this section, the focus is placed on CME2 as it closely impacts Earth. CME2 is characterised using both remote and in situ observations at three different radial distances. With the help of all these pieces of the puzzle, the discrepancy of the magnetic field orientation of the CME throughout the propagation is explored.

%%%---------------------%%%%%%%%%%%%%%%%%%%%%-------------------%%%
\subsection{Deriving the reconnected magnetic flux associated with the CMEs (near $1\;$R$_\odot$)}
\label{sec:mag_reconstruct_sun}

The amount of reconnected magnetic flux is derived using flare-CME statistical relations by previous works as adopted in \citet{Scolini2020}. The relations between the flare peak intensity in soft X-rays and the reconnected flux derived from the flare ribbons and coronal dimmings \citep{Kazachenko2017,Dissauer2018a,Tschernitz2018} are applied. Once the reconnected flux is obtained, the toroidal flux is derived based on the magnetic topology of the CME model. The statistical relations between the reconnected flux, $\phi_r$ (in units of Mx) and the flare peak intensity, $I_{SXR}$ (in units of W~m$^{-2}$) used in this study are as follows: \\
\begin{enumerate}
    \item \citet{Kazachenko2017}: Flare ribbon proxy
    \begin{equation} \label{Kazachenko+2017}
        \text{log}_{10}(\phi_r) = 24.42 + 0.64 \text{log}_{10}(I_{SXR})
    \end{equation}
    
    \item \citet{Dissauer2018a}: Coronal dimming proxy
    \begin{equation} \label{Dissaur+2018}
        \text{log}_{10}(\phi_r) = 23.26 + 0.42 \text{log}_{10}(I_{SXR})
    \end{equation}
    
    \item \citet{Tschernitz2018}: Flare ribbon proxy
    \begin{equation}\label{Tschernitz+2018}
    \text{log}_{10}(\phi_r) = 24.21 + 0.58 \text{log}_{10}(I_{SXR})
    \end{equation}
\end{enumerate}
The values of $\phi_r$ computed from the above relations and their averages are reported for CME1 ($I_{SXR}$=4.5$\times$10$^{-5}\;$W~m$^{-2}$) and CME2 ($I_{SXR}$=1.6$\times$10$^{-4}\;$W~m$^{-2}$) in Table~\ref{tab:cmeflux_obs}. The method for the conversion of $\phi_r$ to toroidal flux ($\phi_t$) for the linear force-free spheromak model \citep[hereafter, referred to as spheromak model;][]{chandrasekhar1958,shiota2016} is followed from \citet{Scolini2019} and yields a value (rounded off to the closest integer) of $\phi_t$=5$\times$10$^{13}\;$Wb for CME1. For the Flux Rope in 3D \citep[FRi3D;][]{Isavnin2016ApJ} model, we follow the FRED method of \citet{Gopalswamy2018} modified for the FRi3D geometry (refer to Appendix~\ref{append:FRi3D_flux}), which results in a total flux (rounded off to the closest integer) of $\phi_{tot} = \phi_t +\phi_p=$1$\times$10$^{14}\;$Wb for CME2.
%{We note that the magnetic flux of the FRi3D model is nearly half of the flux of the spheromak. This depends strongly to the volume, the density and finally, the mass of the two structures, as we explained in Section 5.3.}
{ {We note that the magnetic flux of the CME2 can also be derived by fitting the FRi3D model to the in situ observations (as in Section~\ref{subsec:fri3d_insitu}), and the preference to use this estimate will be described in Section~\ref{subsec:euh_cme_modelling}.}} The choice of employing the spheromak and the FRi3D models for CME1 and CME2, respectively, will be explained in Section~\ref{sec:euhforia}. %The chirality is left-handed, as inferred from the low coronal observations in Section~\ref{subsec:mag_sign_sun}.
\begin{table}
\centering
\begin{tabular}{ p{9cm}| p{1.5cm} p{1.5cm}}
 \hline
 \hline
 {Reconnected flux from statistical relations ($\times 10^{21}$Mx)} &  {CME1} & {CME2} \\
 \hline
 \hline
 \citet{Kazachenko2017} (ribbons) & 4.41 & 9.79 \\
 
 \citet{Dissauer2018a} (coronal dimmings) & 2.74 & 4.63\\
 
 \citet{Tschernitz2018} (ribbons)   & 4.94 &  10.2\\
 
 \hline
 Average & 4.03 & 8.21 \\ %12.09 & 24.62\\
 \hline
 \hline

\end{tabular}
\caption{Reconnected flux using statistical relations observations.}
\label{tab:cmeflux_obs}
\end{table}
%%%---------------------%%%%%%%%%%%%%%%%%%%%%-------------------%%%

\subsection{CME kinematics and geometry in the corona}
\label{subsec:geo_reconstruct}
%https://serpentine-h2020.eu/tools/
On September 9, 2014, Earth was at a longitudinal separation of $167\degree$ and $161\degree$ from STEREO-A and STEREO-B respectively (see Fig.~\ref{fig:sc_relative_pos}). Due to a data gap in STEREO-A during this period, white light coronagraph images from only two viewpoints, \ie \ STEREO-B and LASCO, were used in the reconstruction. STEREO-A did not record data during this time. Although the observations by STEREO-B provided an additional vantage point for the reconstruction, its location on the back side of the Sun made the projected view of the CMEs close to halo CMEs, which made the 3D reconstruction more challenging. %However, it appeared that the CME2 was headed northward off the ecliptic.
The 3D reconstruction of CME1 was performed using the Graduated Cylindrical Shell \citep[GCS;][]{Thernisien2011} model to constrain the geometrical parameters for the spheromak model which will be used in the EUHFORIA simulations in Section~\ref{sec:euhforia}. The GCS parameters (in Stonyhurst coordinates) for CME1 are listed in Table~\ref{tab:gcs_cme1_cme2}: CME latitude ($\theta$), longitude ($\phi$), face-on ($\alpha$) and edge-on ($\delta$) angular half-widths, aspect ratio ($\kappa$=sin$\delta$), and tilt ($\gamma$). The deprojected (3D) speed of the CME leading edge is $v_{3D}$ which is the sum of the radial speed ($v_{rad}$, speed of the CME centre) and the expansion speed ($v_{exp}$, rate of increase of the CME cross-section). The leading edge of the CME is tracked temporally to derive the $v_{3D}$. The spheromak model is launched with $v_{rad} = v_{3D}/(1+\kappa)$ so that the CME cross-section expands self-consistently in the MHD heliospheric domain due to the Lorentz force \citep{Scolini2019}. The CME radius of the spheromak model at $21.5\;$R$_\odot$ is given by 21.5 sin($\alpha$+$\delta$) \ R$_\odot$. The reconstructed images are shown in Fig.~\ref{fig:cme1_3d_fit} in LASCO C3 (top) and STEREO-B (bottom) FOV. As CME2 will be simulated with the FRi3D model (more information later in Section~\ref{sec:euhforia}), its 3D reconstruction is performed with the FRi3D forward modelling tool in order to constrain the parameters appropriately for the simulation. The parameters obtained from the fitting are listed in Table~\ref{tab:gcs_cme1_cme2}: CME latitude ($\theta$), longitude ($\phi$), angular half-width ($\varphi_{hw}$) and half-height ($\varphi_{hh}$), toroidal height ($R_t$), tilt ($\gamma$), flattening ($n$), and pancaking ($\varphi_{p}$). Toroidal speed ($v_{R_t}$) and poloidal speed ($v_{R_p}$) are computed from the temporal fitting of the CME evolution and are similar to $v_{rad}$ and $v_{exp}$ respectively, as mentioned in the context of the spheromak model. The FRi3D model fitted to CME2 in COR-2B and C3 FOV are plotted in Fig.~\ref{fig:cme2_3d_fit}. Using the total speed constrained from the 3D reconstruction and assuming self-similar expansion in the upper corona, the time of injection of the CMEs at 0.1~au EUHFORIA boundary is computed. 

The position of CME2, obtained from the 3D reconstruction, is consistent with the northwestward deflection of the CME and suggests a close-to-flank encounter at Earth if self-similarly extrapolated up to 1~au. Although two viewpoints are reported to improve the reconstruction \citep{VERBEKE2022}, CME1 and CME2 were observed as halo by both LASCO and STEREO-B which could increase the error in the especially critical parameters like speed and half-angle \citep{Kay2020}. % and Fig.~\ref{fig:fri3d_insitu_fit}(b) after being extrapolated beyond 1~au. From this extrapolation, it is evident that CME2 is a close-to-flank encounter at Earth. 
Although the tilt (geometrical inclination) of the fitted flux rope is obtained from this methodology of geometrical reconstruction, the axial magnetic field is ambiguous \ie \ it can be either east-to-west or west-to-east. As it is not straightforward to estimate the vector magnetic field in the middle-to-high corona, we rely on the in situ observations to determine the magnetic field components and hence the flux rope orientation. In the next subsection, in situ observations are used to constrain the CME2 magnetic field orientation at $1\;$au. 

\begin{table}
\centering
\begin{tabular}{ p{3cm} p{1.8cm} | p{3cm} p{1.8cm}}
 \hline
 \hline
 GCS parameters  &  CME1 & FRi3D parameters & CME2\\
 \hline
 \hline
 $\theta$   & 17$\degree$ & $\theta$   & 24$\degree$\\
 $\phi$     & -29$\degree$ & $\phi$     & 15$\degree$\\
 $\alpha$   & 46$\degree$ &  $\varphi_{hw}$  & 50$\degree$ \\
 $\kappa$   & 0.4 &  $\varphi_{hh}$   & 30$\degree$ \\
 $\gamma$   & -55$\degree$ &  $\gamma$   & 45$\degree$\\
 $v_{3D}$   & $696$~km~s$^{-1}$ &  $v_{R_t}$   & $580$~km~s$^{-1}$\\
 $v_{rad}$  & $497$~km~s$^{-1}$ &  $v_{R_p}$  & $363$~km~s$^{-1}$\\
 $r_{spr}$  & 21 R$_\odot$ & &\\
 & &  $n$  & 0.5\\
 & & $\varphi_p$  & 0.5\\
 \hline
 \hline
\end{tabular}
\caption{Parameters from the GCS fitting of CME1 and the FRi3D fitting of CME2.}
\label{tab:gcs_cme1_cme2}
\end{table}

%%%---------------------%%%%%%%%%%%%%%%%%%%%%-------------------%%%
\section{Reconstruction of CME2 from in situ observations at 1~au}
\label{sec:obs_analysis_insitu}
 %In the previous section, the magnetic field orientation of the CMEs was constrained close to $1\;$R$_\odot$. We constrain the magnetic field orientation of the CMEs at 1~au to check for possible rotations during their heliospheric propagation. 
 The in situ magnetic field observations of ME2 from the Earth-bound WIND spacecraft are fit with the FRi3D, Linear Force-Free (LFF) and Circular-Cylindrical (CC) models to derive the chirality and the magnetic axis orientation of the flux rope at $1\;$au. Multiple models are used for validation purposes. {The three selected models have cylindrical or modified-cylindrical configurations and involve the effect of the self-similar expansion of the flux rope, hence making the fittings more realistic \citep[as shown by][]{Vemareddy2016a}.} In addition to constraining and verifying the magnetic field parameters, the motivation behind the in situ fitting is to investigate any rotation between $0.1\;$au and 1~au. As CME1 has no clear rotations in the magnetic field components at $1\;$au, we perform the fittings only for CME2. The comparison of the CME2 orientation near $1\;$R$_\odot$, corona and 1~au is presented at the end of this section. %Results of the fitting are summarised in Table~\ref{tab:cmetilt_obs} and Fig.~\ref{fig:all_insitu_fit}.  

\subsection{FRi3D model}\label{subsec:fri3d_insitu}
The numerical iterative fitting of the FRi3D flux rope to the in situ observations of ME2 is performed using a real-valued version of the genetic algorithm as introduced in \citet{Isavnin2016ApJ}. {{The magnetic field in the FRi3D model is defined by the Lundquist model \citep{Lundquist1950}. The flux rope expansion is implemented by constructing a linearly growing CME cross-section into the model.}} In this model, the tilt parameter provides the latitudinal inclination (positive for counterclockwise and negative for clockwise from the ecliptic on the solar west), and the polarity determines the azimuthal direction (westward is $+1$; eastward is $-1$) of the magnetic field axis. The chirality is negative (positive) for right-handed (left-handed), i.e.\ the opposite of the standard convention. %In addition to the magnetic field parameters and positional coordinates (latitude and longitude of eruption), the angular half-width and half-height also give an idea of the spacecraft crossing through the CME. The projected latitude and longitude of CME launch estimated from 1~au, lie in the range [$-20\degree,0\degree$] and [$10\degree$, $30\degree$] respectively. Combining the information about angular half-width ($\sim50\degree$), half-height ($\sim30\degree$), and tilt ($\sim40\degree$), the flank encounter of the CME at Earth can be inferred.
A detailed description of the parameters can be found in \citet{Maharana2022}. 
%Fig.~\ref{fig:all_insitu_fit}(a) shows the results of FRi3D fitting (in green) that yields a westward left-handed flux rope with a tilt of $+55\degree$. 
The FRi3D fitting (in green), as shown in Fig.~\ref{fig:all_insitu_fit}(a) yields a westward left-handed flux rope with a tilt of $+55\degree$. This fitting also provides an estimate of the total magnetic flux of $0.5\times10^{14}\;$Wb and a twist of $\sim1.5$ associated with CME2 at $1\;$au.
%/home/u0141347/ai.fri3d_events/20140910/trial_weight10_2 : b_25.npy
%chirality [1]
%flattening [0.45463835]
%flux [5.05286762e+13] **avg. sometimes it was 7e13
%half_height [32.32181115]
%half_width [36.20495694]
%latitude [-14.55406355]
%longitude [27.62701283]
%pancaking [0.7485052]
%polarity [0.90211302]
%sigma [2]
%skew [0.]
%tilt [55.59118406]
%toroidal_height [735.91894339   0.95992546]
%twist [1.5] #[1.33579688]
%This model does not give an impact parameter \ie \ the distance of spacecraft crossing from the flux rope axis. Although we have noticed a northward propagation of CME2 in the coronagraph FOV, the FRi3D in situ fits imply negative launch latitude (southward). We speculate that the algorithm could be unreliable in estimating the position of the magnetic ejecta due to the high impact factor of the event. %We can take away the flux rope axis direction, which is north-westward, and its left-handedness. #/home/u0141347/ai.fri3d_events/20140910/trial_weight10_2: chirality [1] flattening [0.45463835] flux [7.05286762e+13] half_height [32.32181115] half_width [36.20495694] latitude [-14.55406355] longitude [27.62701283] pancaking [0.7485052] polarity [0.90211302] sigma [2] skew [0.] tilt [55.59118406] toroidal_height [735.91894339   0.95992546] twist [1.33579688]

\subsection{Linear Force Free model}\label{subsec:lff_insitu}
%The LFF model, by virtue of its simplicity, is not able to fit the magnetic field signatures because of the very characteristics of this magnetic ejecta (\ie expanding structure with a very asymmetric B profile, and slight rotation in the B components especially in the back portion). 
Two fits were obtained by employing the Linear Force Free (LFF) model {{which employs the Lundquist magnetic field configuration \citep{Lundquist1950}}} with two different chiralities ($H$): positive ($H = +1$) and negative ($H = -1$). The results shown in Fig.~\ref{fig:all_insitu_fit}(a) (in red for $H = +1$ and in blue for $H = -1$) consistently agree on the following: (i) the flux rope axis direction ($\theta \sim$25$\degree$, $\phi \sim$-150$\degree$); (ii) the high impact angle (|$z_0$|$\sim$0.88) implying the passage of Earth far away from the axis of the ideal cylindrical structure assumed by the model, and (iii) the low chi-square of both these fits reflect high uncertainty in the fitting. As flux rope chirality remains unchanged during heliospheric propagation \citep{Palmerio2018}, we consider the LFF fitting results with negative chirality based on the source region signatures. %This is consistent with the little rotations observed, and the long duration of the ejecta.

\subsection{Circular-Cylindrical model}\label{subsec:cc_insitu}
\begin{figure}
\centering
    %\subfloat[]{\includegraphics[width=0.4\textwidth]{Figures/20140910_ace_force_free_fit_posH.png}}
    %\subfloat[]{\includegraphics[width=0.4\textwidth]{Figures/20140910_ace_force_free_fit_negH.png}} \\ 
    %\subfloat[]{\includegraphics[width=0.4\textwidth]{Figures/fig_56.png}}
    \subfloat[]{\includegraphics[width=0.5\textwidth,trim={0cm 1.0cm 0cm 0cm},clip=]{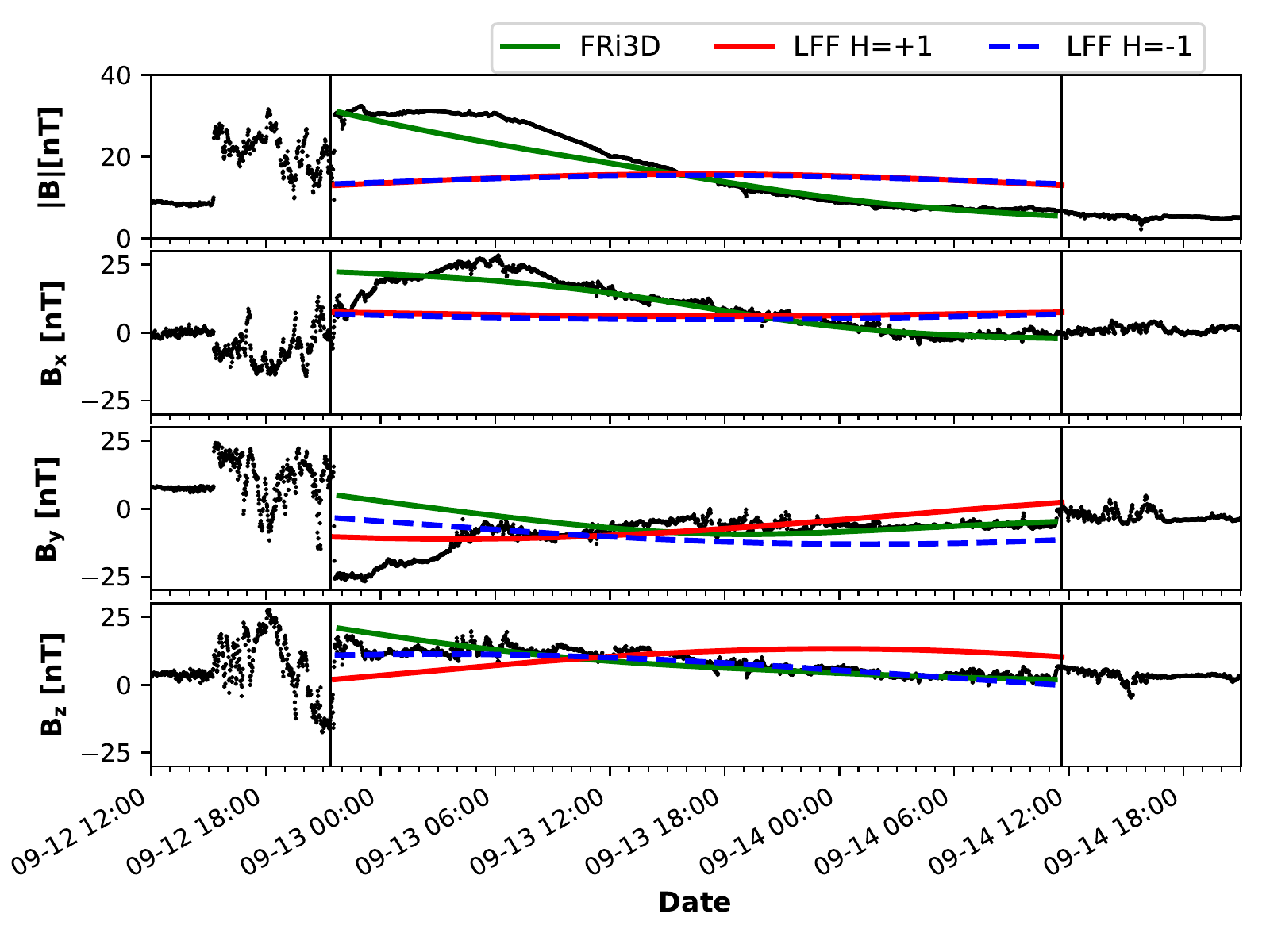}}\\
    \subfloat[]{\includegraphics[width=0.5\textwidth,trim={0cm 3cm 1.3cm 1.7cm},clip=]{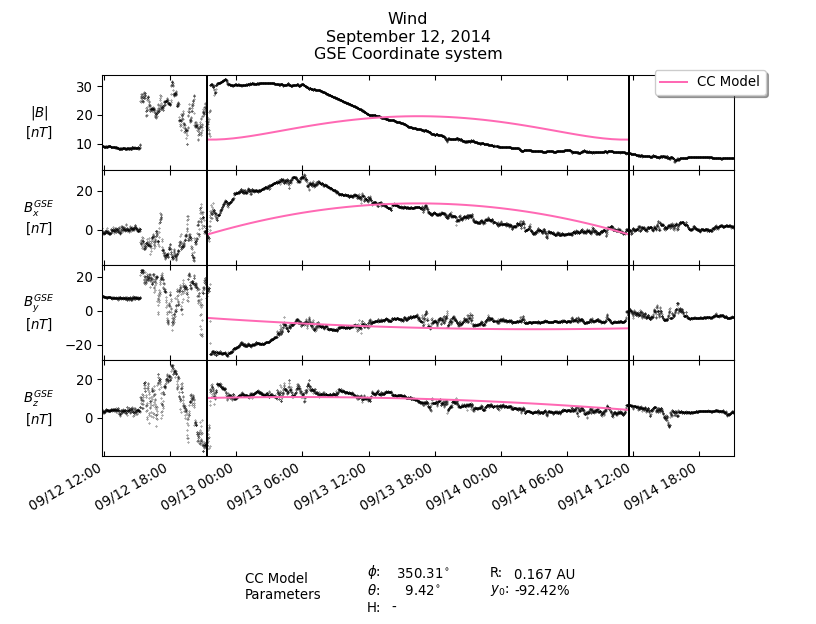}}
    %\caption{Fitting of in situ observations of ME2 at 1~au with various models. (a) Linear force-free (LFF) fit with right-handed chirality, and (b) left-handed chirality. The fitted parameters of the model are listed at the top of the images. (c) FRi3D model fitting to the magnetic field components and the speed; (d) Circular-cylindrical (CC) fit. The fitted parameters of the model are listed at the bottom of the image \citep{Nieves2019}.}
    \caption{Fitting of in situ observations of ME2 at 1~au with various models: (a) FRi3D model (green), LFF fit with right-handed chirality (red), and LFF fit with left-handed chirality (blue); (b) CC fit adapted from the WIND ICME catalogue \citep[][]{Nieves2018}. The vertical lines in black in both plots correspond to the ME2 boundary as per the same catalogue. The fitted parameters of the model are discussed in Section~\ref{subsec:fri3d_insitu}, \ref{subsec:lff_insitu},  and \ref{subsec:cc_insitu}.}
    \label{fig:all_insitu_fit}
\end{figure}

The analysis of the in situ signatures with the circular-cylindrical analytical flux rope model \citep[CC model;][]{Nieves2016} is adapted from the WIND ICME catalogue \citep[\url{https://wind.nasa.gov/ICME_catalog/ICME_catalog_viewer.php};][]{Nieves2018} (see Fig.~\ref{fig:all_insitu_fit}(d)). {{The magnetic field configuration of this model is based on the non-force-free approach by relating the magnetic field vector with its current density, as proposed by \citet{Hidalgo2002a}. In this model, the cross-section distortion, expansion, curvature and deformation are implemented following \citet{Hidalgo2002b} to reconcile CME with ICME from in situ, remote sensing and MHD simulations perspectives.}}. The orientation of the flux rope quantified by the latitude, $\theta=9\degree$ (positive) and the longitude, $\phi=350\degree$ ($>180\degree$) corresponds to a low-inclined northwestward flux rope. Negative helicity suggests the left-handedness of the ME. %$R$ is the radius of the flux rope cross-section. 
Impact parameter, $|y_0|\sim0.9R$ %or $y_0 = -92\%$
implies that the smallest distance of the spacecraft to the flux rope axis ($y_0$) relative to the flux rope radius (R), is almost at the edge of the flux rope boundary.  \\

The conclusions from the different in situ reconstruction techniques are: (a) the axial magnetic field of ME2 has a northwest orientation at $1\;$au, (b) the flux rope is left-handed, and (c) a high impact parameter implies flank encounter. It must be noted that the uncertainty associated with determining the exact flux rope orientation increases in cases with high-impact parameters as compared to the head-on impacts \citep{Riley2004,Al-Haddad2013}. The conclusions (a) and (b) suggest the magnetic topology of CME2 to be NWS \citep{Bothmer1998} at 1~au, contrary to SEN as inferred close to $1\;$R$_\odot$ in Section~\ref{subsec:mag_sign_sun}. %However, we do note that the error associated with such fits increases in the case of high-impact parameter cases such as the one being analysed.
%%%---------------------%%%%%%%%%%%%%%%%%%%%%-------------------%%%
\subsection{Discrepancy in CME2 orientation from observations at different locations}
%%%---------------------%%%%%%%%%%%%%%%%%%%%%-------------------%%%

\begin{figure}
    \centering
    \includegraphics[width=0.6\textwidth]{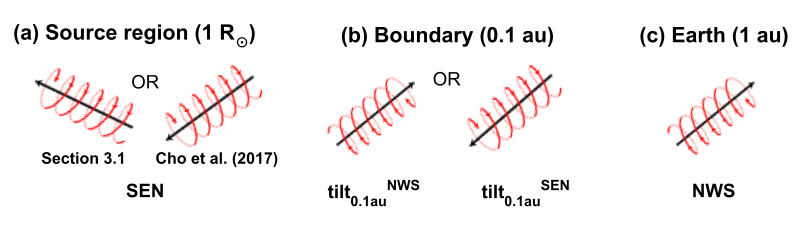} %width = 0.6 --> tilt_explanation2.png
    \caption{{{Schematic representation of the}} CME2 orientation inferred from different observational proxies at different locations - {{(a) Close to 1~$R_\odot$, based on the analysis of the source region in Section~\ref{subsec:mag_sign_sun} and the analysis of \citet{Cho2017}; (b) Close to the 0.1~au, based on the 3D reconstruction of the white-light images (Section~\ref{subsec:geo_reconstruct}); (c) At 1~au, based on the in situ observations (Section~\ref{sec:obs_analysis_insitu}).}}}
    \label{fig:tilt_explanation1}
\end{figure}

The analysis of the flux rope during the eruption in Section~\ref{subsec:mag_sign_sun} suggests an SEN topology. However, an analysis of the in situ signatures of the magnetic ejecta at 1~au points to an NWS topology. This discrepancy in the flux rope orientation retrieved for CME2 at the Sun and 1~au is the focus of our investigation in the following sections. %This could support the hypothesis of a heavy rotation of the flux rope in the low corona. %As per \citet{Cho2017}, the PIL orientation of the full-halo CME is estimated to be $245\degree$ (southeast-ward). %But given the possibility of extreme rotation, we speculate the CME could have rotated $\sim$270$\degree$.
As per the PIL orientation of the CME2 source (Fig.~\ref{fig:AR_wv_hmi}(b)), the flux rope orientation can be approximated to be SEN pointing in the northeast direction. However, the details of the eruption from \citet{Dudik2016} and \citet{Vemareddy2016} suggest a shearing rotation and deflection motion that could have led to the eruption of a southeast pointing SEN flux rope. {{The two possible cases are depicted in Fig.~\ref{fig:tilt_explanation1}(a)}}. \citet{Cho2017} also estimate the PIL orientation of the CME2 to be southeast SEN during the eruption. Close to 0.1~au, CME2 can have two axial magnetic field directions for the same geometrical tilt as shown in Fig.~\ref{fig:tilt_explanation1}(b) - either SEN or NWS. We propose two possible scenarios henceforth. 
First, assuming a dominant low coronal rotation and no significant rotation in the inner heliosphere (0.1~au to 1~au), the northwestward directed tilt constrained close to 0.1~au (hereafter, tilt$^{NWS}_{0.1au}$) turns out to be consistent with the tilt constrained at 1~au (hereafter, tilt$_{1au}$ as in Fig.~\ref{fig:tilt_explanation1}(c)). The repercussion of this assumption is a physical rotation of CME2 by $\sim$180$\degree$ - $270\degree$ (assuming uncertainties in defining the PIL) in an anti-clockwise direction owing to the left-handed chirality of the CME \citep{Green2007,Lynch2009}. \citet{Vemareddy2016} suggest the eruption of CME2 was triggered by a helical kink instability driven by sunspot rotation. Such kink-unstable magnetic flux ropes are known to produce CME rotation in the low corona by converting their twist into writhe \citep{Kliem2012} which could have resulted in such a significant rotation of CME2. The second scenario suggests a partial anti-clockwise rotation in the low corona leading to the southeast SEN flux rope at 0.1~au (tilt$^{SEN}_{0.1au}$), followed by an additional rotation in the heliosphere by about $\sim$180$ - $270$\degree$ to reach the reconstructed tilt$_{1au}$ at 1~au. The second scenario seems less probable as previous studies suggest that most CMEs cease to undergo significant rotation and deflection at larger heliospheric distances and instead propagate self-similarly further away from the Sun \citep{demoulin2009,Isavnin2014,balmaceda2020}. We investigate the possibility of these two scenarios with numerical simulations in the next section.   

%%%---------------------%%%%%%%%%%%%%%%%%%%%%-------------------%%%
\section{MHD modelling with EUHFORIA}
\label{sec:euhforia}
%%%---------------------%%%%%%%%%%%%%%%%%%%%%-------------------%%%
In this section, we present the simulation setup of the heliospheric propagation of the CMEs using the physics-based MHD model EUropean Heliospheric FORecasting Information Asset \citep[EUHFORIA]{Pomoell2018}. The aim is to match the observations at 1~au measured by the WIND spacecraft. The questions we seek to answer are: (a) What is the orientation of CME2 at 0.1~au that must be injected to obtain the correct signature of ME2? (b) What is the role of CME1 in forming the magnetic field rotation in the sheath region of CME2? 

\subsection{EUHFORIA setup}
\label{subsec:euhforia_setup}
EUHFORIA consists of two parts: a coronal domain and a heliospheric domain. The coronal part is a 3D semi-empirical model based on the Wang-Sheeley-Arge (WSA, \citet{Arge2004}) model, which provides the solar wind plasma conditions at the inner boundary of EUHFORIA, \ie \ 0.1~au. It is driven by the photospheric magnetic field via synoptic magnetogram maps. % from Global Oscillation Network Group (GONG) of National Solar Observatory or the Mount Wilson Observatory. 
More details about the coronal model can be found in \citet{Pomoell2018} and \citet{Asvestari2019}. The heliospheric part is a 3D time-dependent model of the inner heliosphere that numerically solves the ideal MHD equations, including gravity, using a cell-average finite volume method in the Heliocentric Earth EQuatorial (HEEQ) coordinate system. The constrained transport approach is applied to advance the magnetic field components in a divergence-free way. The boundary conditions at 0.1~au of this part are obtained from the coronal model. The computational domain extends from 0.1~au to 2~au in the radial direction, $\pm80\degree$ in the latitudinal direction, and 0-360$\degree$ in the longitudinal direction. EUHFORIA enables the injection of the CMEs at the inner boundary as time-dependent boundary conditions which are then self-consistently evolved by MHD equations. There are three functional CME models: (1) the cone model \citep{Pomoell2018}: a simplified non-magnetised spherical blob of plasma; (2) the LFF spheromak model \citep{Verbeke2019}: an improvement over the cone model by the inclusion of an internal magnetic field configuration; and (3) the FRi3D model \citep{Maharana2022}: an upgrade over the spherical shape of the spheromak model for improving the modelling of flank encounters and deformations. EUHFORIA version 2.0 has been used for the simulations in this work. The radial resolution of the computational mesh is 0.0074~au (corresponding to $1.596\;$R$_{\odot}$) for 256 cells in the radial direction, and the angular resolution is $4^\circ$ in the latitudinal and $2^\circ$ in the longitudinal directions, respectively. 

\subsection{The background solar wind}
\begin{figure}
    \centering
    \includegraphics[width=0.5\textwidth]{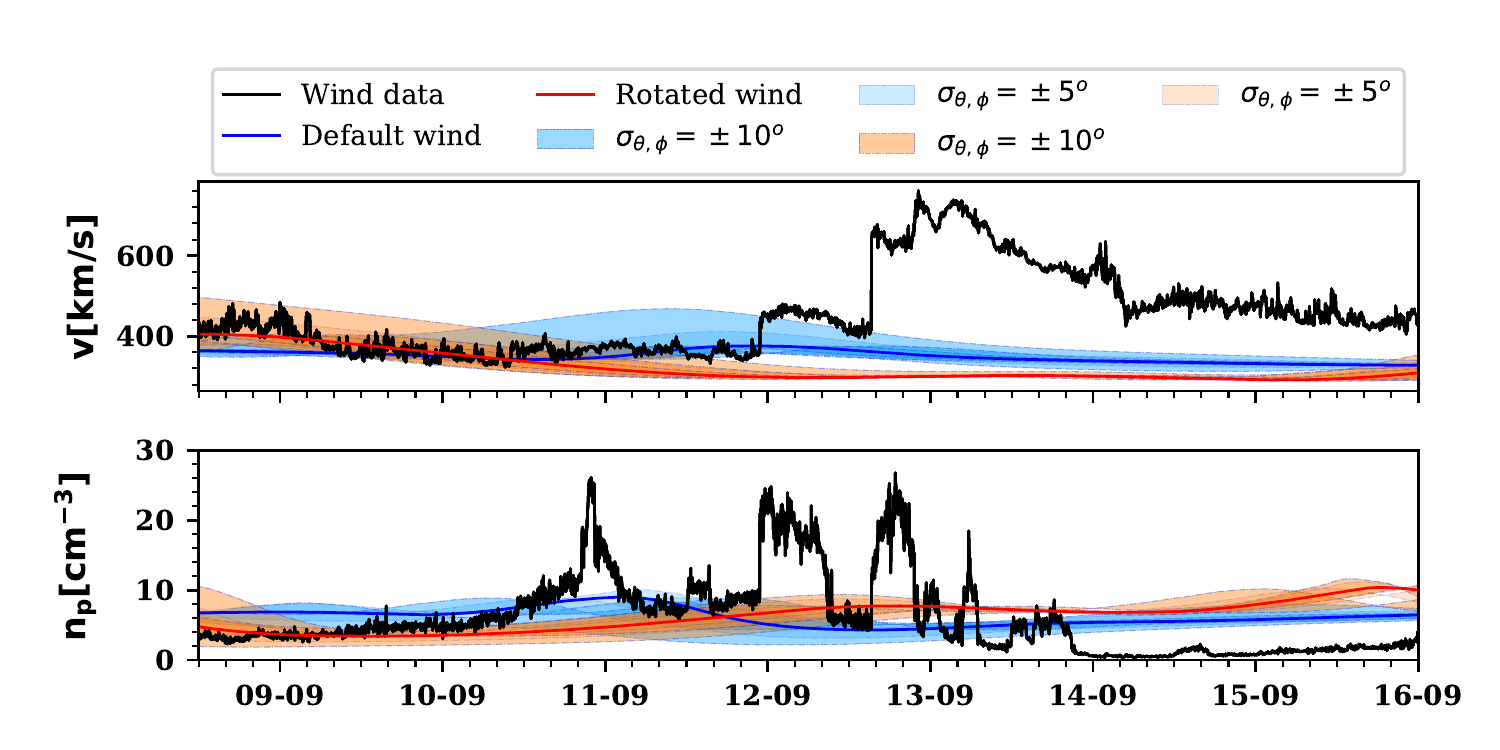}
    \caption{Background solar wind modelled with the default EUHFORIA coronal model setup using the synoptic magnetogram from GONG on September 8, 2014, at 23:00 (in blue), and the rotated solar wind (in red). The shaded regions provide an error estimate in $\pm$5-10$\degree$ in latitude ($\sigma_\theta$) and longitude ($\sigma_\phi$) around Earth. Corresponding WIND observations are plotted in black.}
    \label{fig:sim_wind}
\end{figure}

We perform the first EUHFORIA simulation by evolving the solar wind as a boundary condition without the insertion of CMEs to obtain an optimal ambient medium with reasonable plasma properties in which CMEs can propagate. The background solar wind is modelled using the synoptic magnetogram from Global Oscillation Network Group (GONG) on September 8, 2014, at 23:00~UT (\text{mrbqs140908t2314c2154$\_$055.fits.gz}). With this magnetogram fed as the boundary condition to the default coronal model of EUHFORIA, we obtained a high-speed stream traversing through Earth with its peak reaching $\sim 3\;$days later as compared to the in situ observations. Hence, we rotated the inner boundary map of extrapolated solar wind plasma and magnetic field properties by 40$\degree$ westward, in order to make the high-speed stream arrive earlier at Earth, and reproduce more accurately the actual CME propagation and its position with respect to the high-speed stream. %so that it does not bias the propagation of the CMEs in the simulations. 
The speed and the proton number density profiles from both simulations, the default wind (in blue) and the rotated wind (in red) are shown in Fig.~\ref{fig:sim_wind}. %Speed $\sim$400~km~s$^{-1}$ on September 9, 2014. 

\subsection{Modelling of CMEs}
\label{subsec:euh_cme_modelling}
In this work, we employ the spheromak model and the FRi3D model to simulate CME1 and CME2, respectively. We refrain from modelling both CMEs with FRi3D due to an implementation limitation affecting the injection of consecutive FRi3D CMEs into the heliospheric domain. As the legs of a first CME simulated with FRi3D would remain connected to the inner boundary, the insertion of a second CME would raise numerical complications. As the main focus of this study is CME2, FRi3D is used for improved modelling of the magnetic field components and the spheromak model is used for CME1. %FRi3D has a global extended CME geometry which is capable of modelling the CME flank encounters which can be missed by simplistic spherical models like the cone or the spheromak model. 
{{We first experimented by using the spheromak model for simulating both CMEs. When CME2 was modelled with spheromak, the CME had to be launched almost along the Sun-Earth line (although the real CME2 event was a flank encounter) in order to model its interaction with CME1. This is because of the inability of the spheromak model to reproduce the flank impact of CME2 due to the lack of CME legs in the model. Recent studies point to additional drawbacks of using the spheromak model in modelling the interplanetary propagation of the CMEs. The magnetic moment of the spheromak model tends to tilt and align itself with the ambient solar wind magnetic field. As the spheromak model is not anchored to the Sun unlike a real CME, it is free to undergo unreal rotation in the heliosphere due to the spheromak tilting instability \citep{Asvestari2022}. Therefore, it will be unreliable for understanding the possibility of the actual rotation of the CMEs in interplanetary space. We had to adjust the density of the spheromak model in an ad hoc manner to limit the inherent spheromak tilting. This resulted in an overestimated number density profile of the CME2, and yet the magnetic field profiles of CME2 were not appropriately reproduced. %In order to avoid too much tampering with the CME input parameters without a systematic rationale, we preferred to use the FRi3D model for the main event, CME2. 
Due to the numerical constraint of using the FRi3D model to simulate both the CMEs consecutively, we chose to use the spheromak model for CME1 and shifted it towards the Sun-Earth line in a calculated manner to represent the leg of CME1.}}
%Hence, we use the possibility of applying two different magnetised CME models for simulating multiple CMEs in the same EUHFORIA simulation. 
We constrain the CME input parameters for the spheromak and the FRi3D model following the methods described by \citet{Verbeke2019}, \citet{Scolini2019}, and \citet{Maharana2022}. The geometrical and kinematic parameters are obtained from the 3D reconstruction of the CMEs in the solar corona as detailed in Section~\ref{subsec:geo_reconstruct}. The magnetic field parameters for the CME models are constrained using observations of the source region as detailed in Section~\ref{sec:mag_reconstruct_sun}. All the parameters are summarised in Table~\ref{tab:syn_euh_params}.

\begin{table}
\centering
\begin{tabular}{  p{2.0cm} | p{2.5cm} p{2.5cm} }
 \hline
 \hline
 Simulations & CME1 & CME2 \\
 \hline
 \hline
 Run1 & - & FRi3D~(tilt$_{0.1au}^{SEN}$) (2014-09-12 18:13~UT) \\
 \hline
 Run2 & - & FRi3D~(tilt$_{0.1au}^{NWS}$) (2014-09-12 18:23~UT) \\
 \hline
 Run3 & Spheromak (2014-09-11 23:23~UT) & - \\
 \hline
 Run4 & Spheromak (2014-09-11 23:23~UT) & FRi3D~(tilt$_{0.1au}^{NWS}$) (2014-09-12 12:33~UT) \\
 \hline
 \hline
 Observed ToA & 2014-09-11 22:50~UT &  2014-09-12 15:17~UT\\
 \hline
 \hline
\end{tabular}
\caption{List of EUHFORIA simulations, the CME models used and the time of arrival of the shocks (ToA; datetime in yyyy-mm-dd HH:MM format) of the CMEs at Earth in the EUHFORIA simulations. The observed ToA of the CMEs from the Wind ICME catalogue is provided for comparison.}
\label{tab:cme_euh_list}
\end{table}

{{We perform three numerical experiments (Run1, Run2, and Run3) involving one CME (CME1 or CME2) each time to determine the parameters of the individual CMEs at the heliospheric boundary that match the observations at 1~au. The results of these experiments are used to perform the final simulation (Run4) involving both the CMEs.}} %perform four simulations in total. 
The first two simulations labelled Run1 and Run2, aim to investigate the orientation of CME2 that must be injected at 0.1~au to reproduce the magnetic field components when propagated to 1~au, before introducing CME1 ahead of it. All the parameters corresponding to the FRi3D model are kept the same in these two runs except for the polarity. Run1 is initialised with the south-eastward tilt$_{0.1au}^{SEN}$ and Run2 with north-westward directed tilt$_{0.1au}^{NWS}$. The magnetic flux value used for the FRi3D model in Run1 and Run2 is $0.5\cdot 10^{14}\;$Wb which is half the value constrained near {{the photospheric surface ($1\;$R$_\odot$) in Section~\ref{sec:obs_analysis_remote}. This value is not ad hoc, but is consistent with the total magnetic flux value constrained from the fitting of the FRi3D model to the in situ observations at 1~au in Section~\ref{subsec:fri3d_insitu}}}. \citet{Maharana2022} showed that the FRi3D model expands faster and arrives earlier when initialised with the magnetic flux value constrained using the methodology involving the remote-sensing observations, and hence can be initialised with a lower flux {{estimated from in situ observations for better prediction accuracy}}. In Run3, only CME1 is simulated with the spheromak model to assess its independent signature at Earth. The fourth simulation, Run4, is aimed to investigate the effect of CME1 in preconditioning the propagation of CME2, and the consequence of CME-CME interaction on the geoeffectiveness of the impact at Earth. Run4 will {{use the input parameters of}} CME1 from Run3, and {{that of}} CME2 from the best simulation between Run1 and Run2 based on the results of Section~\ref{subsec:cme2_propagation}. As the geometrical reconstruction points to a glancing blow of CME1 at Earth and the spheromak model does not possess legs, it is possible to miss the effect of its flank encounter on the interaction with CME2. Hence, we shift CME1 $\sim 15\degree$ westward in longitude and $\sim 5\degree$ northward in latitude in order to better reproduce the effect of its legs in Run4. In the simulation domain, we place virtual spacecraft around Earth separated by an angular distance of 5$\degree$ and 10$\degree$ in latitude and longitude to capture the variability of the results in the vicinity of Earth. Additional virtual spacecraft are placed along the Sun-Earth line with a radial separation of 0.1~au. {{The standard mass densities used for the spheromak and the FRi3D model are $10^{-18}$~kg~m$^{-3}$ and $10^{-17}$~kg~m$^{-3}$, respectively. According to \citet{Maharana2022}, the typical volume of the flux rope geometry (as in the case of FRi3D) requires a higher standard density in the order of $10^{-17}$~kg~m$^{-3}$ (supported by observations in \citet{Temmer2021}) to enhance the modelling accuracy of mass of a CME modelled with FRi3D at $0.1$~au. However, the spherical volume of the spheromak model is up to $2$-$3$ orders of magnitude higher than that of the FRi3D model and hence, a comparable mass can be modelled a density of $10^{-18}$~kg~m$^{-3}$. Therefore, CME1 and CME2 have different initial mass densities depending on the CME model used.}}

\begin{table}
\centering
\begin{tabular}{ p{4cm}||p{2.5cm}|p{2.5cm}}
 \hline
 \hline
 \multicolumn{3}{c}{{Input parameters}} \\
 \hline
\hline
    &  CME1 & CME2 \\
 \hline
 CME model & Spheromak & FRi3D \\
  \hline
  \multicolumn{3}{c}{Geometrical} \\
 \hline
 Insertion time   & 2014-09-09~04:24~UT & 2014-09-10~20:14~UT \\
 Speed   & $450$~km~s$^{-1}$ &  $500$~km~s$^{-1}$\\
 Latitude    & $22\degree$ & $24\degree$ \\
 Longitude   & $-14\degree$ & $15\degree$ \\
 Half-width  & - & $50\degree$ \\
 Half-height & - & $30\degree$ \\
 Radius      & $21$~R$_\odot$ & - \\
 Toroidal height & - & $13.6$~R$_{\odot}$\\
\hline
 \multicolumn{3}{c}{{Magnetic field}} \\
 \hline
 Chirality   & $-1$ & $+1$$^*$ \\ %\footnote{Right-handedness in FRi3D is defined as $-1$.}

 Polarity    & - & $+1$($-1$) \\
 Tilt        & $-135\degree$ $^{**}$ & $45\degree$ \\
 Toroidal magnetic flux & $5\cdot 10^{13}$~Wb & - \\
 Total magnetic flux & - & $5\cdot10^{13}$~Wb \\
 Twist       & - & $1.5$\\
 \hline
 \multicolumn{3}{c}{{Deformation}} \\
 \hline
 Flattening  & - & $0.5$\\
 Pancaking   & - & $0.5$\\

\hline
 \multicolumn{3}{c}{{Plasma parameters}} \\
\hline
 Mass density     & $10^{-18}$~kg~m$^{-3}$ & $10^{-17}$~kg~m$^{-3}$\\
 Temperature & $0.8 \cdot 10^6$~K & $0.8 \cdot 10^6$~K \\
 \hline
 \hline
\end{tabular}
\caption{CME parameters used in the EUHFORIA simulations employing the spheromak model for CME1 in Run3 and Run4, and the FRi3D model for CME2 in Run1, Run2 and Run4. The only change for Run1 is in the polarity parameter of the FRi3D model, which is -1 (eastward) as opposed to +1 (westward) for Run2 and Run4.  \\
$^*$FRi3D chirality is implemented with an opposite convention \ie \ -1 for right-handedness and +1 for left-handedness. \\
{{$^{**}$Conventionally, a left-handed tilt of $0\degree$ configuration of the spheromak model is a westward flux-rope normal to the Sun-Earth line. Hence, the eastward tilt of $45\degree$, the spheromak model is rotated \textit{anti-clockwise} by $135\degree$, i.e., $-135\degree$.}}}
\label{tab:syn_euh_params}
\end{table}

%the energetic ions, the early CME propagation and the flux intensity during the period of September 9, 2014 and September 12, 2014

\section{Simulation results and discussion}
\label{sec:results}
A summary of the EUHFORIA simulations performed in this study, including the CME models used for each CME, and the time of arrival of the corresponding CME shock (ToA, datetime in yyyy-mm-dd HH:MM format) of the CMEs in the simulations, is provided in Table~\ref{tab:cme_euh_list}.
%%%---------------------%%%%%%%%%%%%%%%%%%%%%-------------------%%%
\subsection{Propagation of CME2 only}
\label{subsec:cme2_propagation}
%%%---------------------%%%%%%%%%%%%%%%%%%%%%-------------------%%%
 
\begin{figure}
    \centering
    \includegraphics[width=0.5\textwidth]{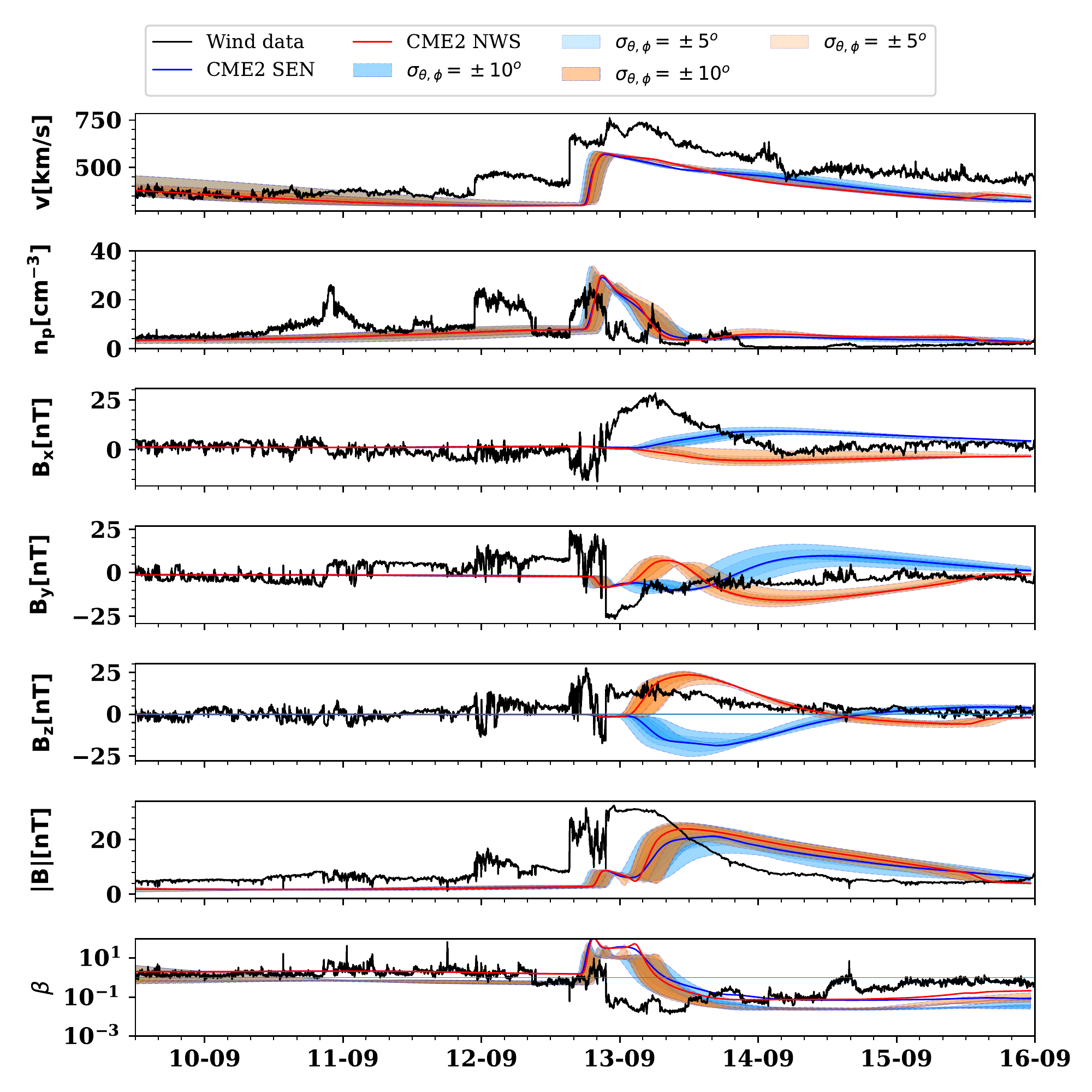}%{Figures/Earth_20140910_newGCS_wind7_fri+45_p+1_lat24_lon15_0p5e14_v500_tw1p5_20140910_newGCS_wind7_fri+45_p-1_lat24_lon15_0p5e14_v500_tw1p5.png}%1_lat24_lon15_0p5e14_v500_tw1p5_20140910_newGCS_wind7_fri+45_p+1_lat24_lon15_0p5e14_v500_tw1p5.png}%20140910_newGCS_wind2_fri+65_p+1_20140910_newGCS_wind2_fri-25_p-1_Earth.png}
    \caption{Time series plot showing the comparison between the simulations with different orientations of CME2 modelled with FRi3D, at Earth - Run1 (blue) is simulated with tilt$_{0.1au}^{SEN}$ {{(deduced based on the orientation close to 1~R$_\odot$)}} and Run2 (red) with tilt$_{0.1au}^{NWS}$ {{(similar to the tilt at 1 au)}}. {{Both the orientations follow the same geometrical tilt derived from the 3D reconstruction at 0.1~au but with two different axial magnetic field orientations.}} From top to bottom: speed (v), proton number density ($n_p$), $B_x$, $B_y$, $B_z$, magnetic field strength (|$B$|) and plasma beta ($\beta$). The solid line and the shaded regions show the profile at Earth and in the 5-10$\degree$ latitudinal and longitudinal offset around Earth respectively. }
    \label{fig:fri+65+1_fri-25-1}
\end{figure}
The comparison of Run1 and Run2 is shown in Fig.~\ref{fig:fri+65+1_fri-25-1}. Speed and density are modelled similarly in both cases. The arrival time of CME2 in both Run1 and Run2 is delayed by $\sim3\;$hours as compared to the observed ToA of the shock. Using tilt$^{NWS}_{0.1au}$ in Run2, the prolonged positive $B_z$ component is well reproduced and the negative $B_y$ component better matches observations compared to the use of tilt$^{SEN}_{0.1au}$ in Run1. Through this experiment, we have developed an understanding of the circumstances that led to the formation of the positive $B_z$ component in ME2 instead of the predicted prolonged negative $B_z$ component. Run1 (tilt$^{SEN}_{0.1au}$) does not seem to rotate significantly in the heliospheric domain of our simulation to match the observations at $1\;$au (tilt$^{NWS}_{0.1au}$). This suggests that the flux rope must have undergone rotation in the corona (i.e. within $0.1\;$au) up to reaching tilt$^{NWS}_{0.1au}$, and it then would have propagated in the heliosphere without significant rotation resulting in the observed magnetic field profile at $1\;$au. %Assuming that rotation (if any) happens self-consistently and similarly, for both Run1 and Run2 in the heliospheric part, low corona rotation seems like a necessary condition in order to obtain the observed magnetic field profile at 1~au. %Although it might be argued that the negative $B_z$ signature in the CME2 front in Run1 could be instrumental in the formation of the negative $B_z$ in the sheath, the sign of the $B_y$ component is not consistent.

%%%---------------------%%%%%%%%%%%%%%%%%%%%%-------------------%%%
\subsection{Propagation of CME1 only}
\label{subsec:cme1_propagation}
\begin{figure}
    \centering
    \includegraphics[width=0.5\textwidth]{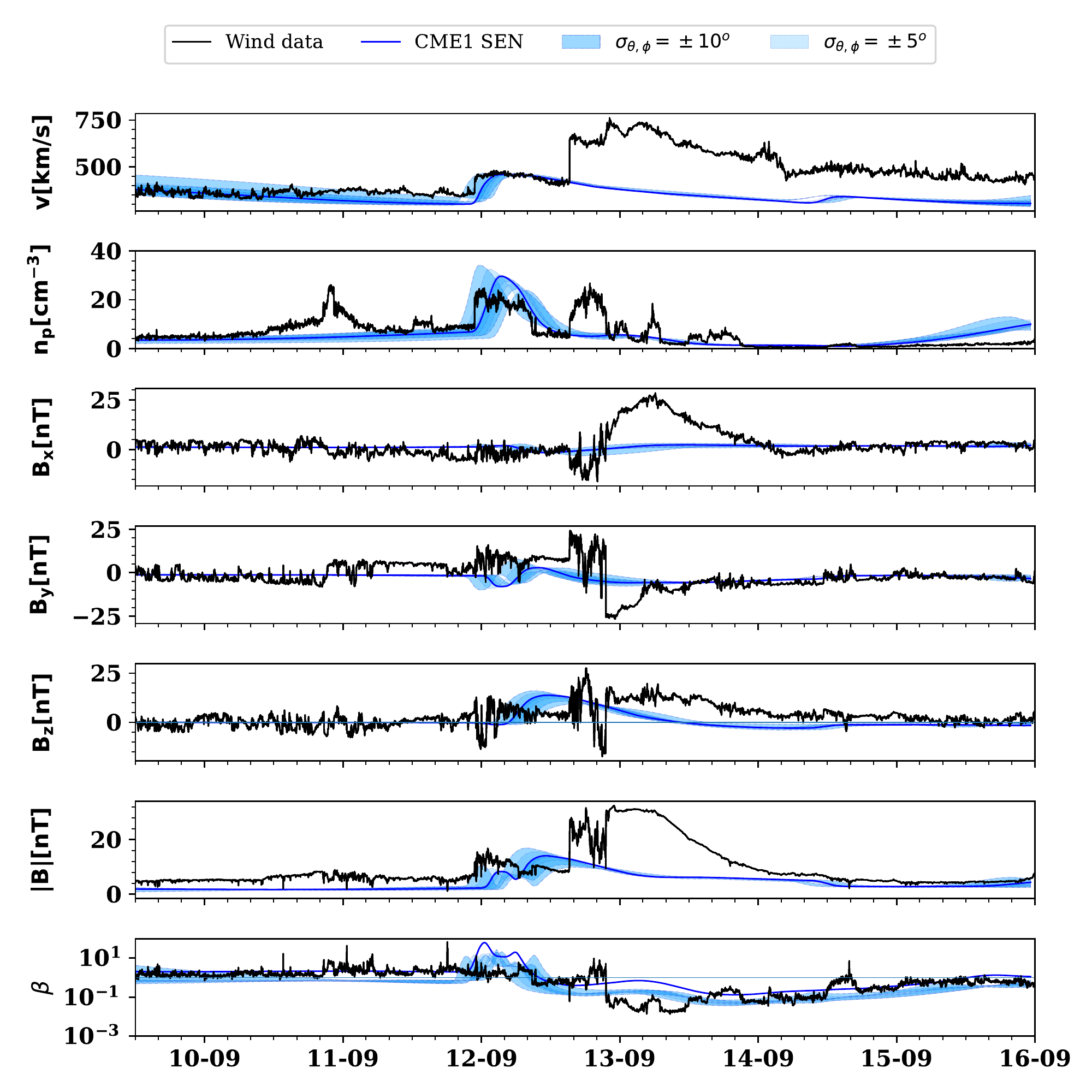} %{Figures/Earth_20140910_newGCS_wind7_s-135_v450_lon-14.png} %
    \caption{Results of Run3 where CME1 is modelled with the spheromak model. The figure description is similar to that of Fig.~\ref{fig:fri+65+1_fri-25-1}.}
    \label{fig:sim_cme1_only}
\end{figure}

%%%---------------------%%%%%%%%%%%%%%%%%%%%%-------------------%%%
This section discusses the results of Run3. We first initialised the CME with $v_{rad}$=497~km~s$^{-1}$, and yet, the CME arrived at Earth $\sim4$ hours earlier than the observed time of arrival. As the main purpose of this study is to understand the CME magnetic field signatures rather than predicting the CME arrival times, we optimised the $v_{rad}$ to $450\;$km~s$^{-1}$ in order to model the latter interaction with CME2 in Run4 more accurately. The modelled time series of the physical parameters at Earth are presented in Fig.~\ref{fig:sim_cme1_only}. After reducing the CME speed as described above, the shock of CME1 is modelled to arrive on 2014-09-11~23:23~UT (43 minutes later than the actual CME1 arrival), as visible in the sharp increases in the speed and proton density profiles. The drop in $\beta$ at $\sim $2014-09-12~12:00~UT is due to the passage of ME1 at its flank i.e., the southwest portion of the magnetic ejecta associated with CME1 (see figures in Section~\ref{subsec:cme1+cme2_propagation}). There is no clear rotation in the magnetic field components. A leading positive $B_z$ signature is obtained (slightly overestimated) followed by a weak trailing negative $B_z$ component.  

%%%---------------------%%%%%%%%%%%%%%%%%%%%%-------------------%%%
\subsection{Propagation of CME1 followed by CME2}
\label{subsec:cme1+cme2_propagation}
\begin{figure}
    \centering
    \includegraphics[width=0.5\textwidth]{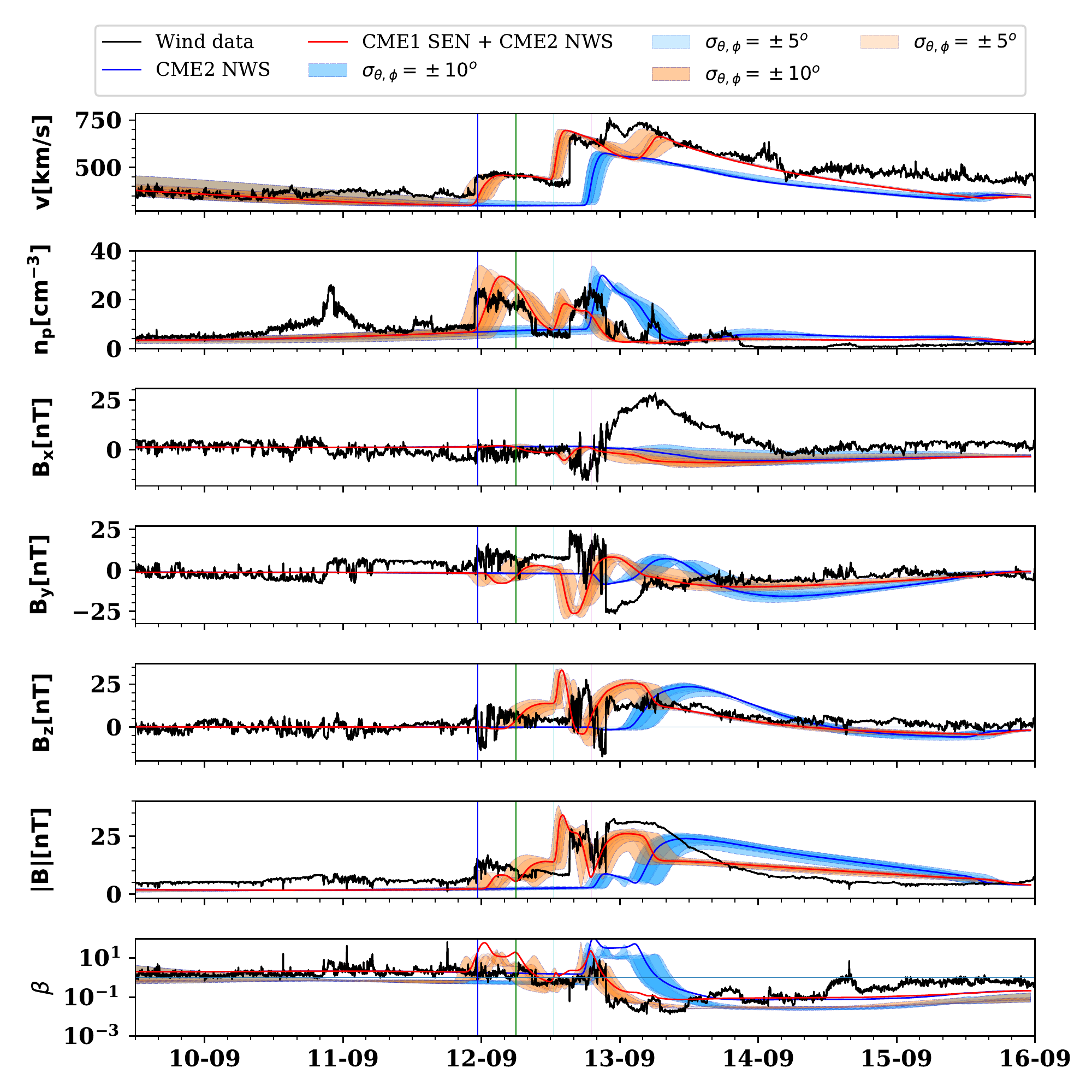}%newGCS_wind7_fri+45_p+1_lat24_lon15_0p5e14_v500_tw1p5_newGCS_wind7_s-135_v450_lon-14_fri+45_p+1_lat24_lon15_0p5e14_v500_tw1p5.png}%{Figures/20140910_newGCS_wind2_fri+65_p+1_Earth_WIND.png}
    \caption{Results of Run2 (blue; CME2 NWS) and Run4 (red; CME1 SEN + CME2 NWS) are compared. CME1 and CME2 parameters in Run4 correspond to Run3 and Run2 respectively. Figure description is similar to Fig.~\ref{fig:fri+65+1_fri-25-1}. S1, the start of ME1, S2 and the start of ME2 are marked with blue, green, cyan and magenta vertical lines respectively in Run4.}
    \label{fig:sim_cme1_cme2}
\end{figure}

%%%---------------------%%%%%%%%%%%%%%%%%%%%%-------------------%%%
%% Plasma properties of the CME shock mainly - arrival time and density
Based on the previous runs, we find that the most accurate modelling of the CME2 signatures at 1~au (in Run4) arises from the simulation of CME1 as in Run3 and of CME2 as in Run2. The results of Run4 are discussed and compared to Run2 in this section. In Fig.~\ref{fig:sim_cme1_cme2}, plasma and magnetic field properties of Run4 are over-plotted on the results of Run2 to distinguish the features due to the possible CME-CME interaction in the presence of CME1 in Run4. The solid vertical lines in blue, green, cyan and magenta correspond to the S1, the start of ME1, S2 and the start of ME2 in Run4. The shaded regions around the solid line of the simulation time series represent the same physical properties in $\pm$5-10$\degree$ vicinity in the latitude ($\sigma_{\theta}$) and longitude ($\sigma_{\phi}$) around Earth. We first analyse the speed and number density in the time series plot at Earth to compare the effect of CME1 in Run4. In the absence of CME1 (Run2), CME2 arrived at Earth $\sim 3\;$hours later than observations. While, in the presence of CME1 (Run4), CME2 arrived at Earth $\sim 3\;$hours earlier with respect to the observations, hence implying that it was sped up by $\sim 6\;$hours by CME1. The passage of CME1 creates a low-density region ahead of CME2 which lets it expand faster, leading to a higher shock speed and a depletion in density inside the flux rope (the $n_p$ peak is lower in Run4 than Run2). {{In Run4, the difference between the arrival time of the CMEs is $\sim$13~hours. It is to be noted that ToA is the arrival time of the CME shock. The sheath and the magnetic ejecta following the shocks are extended structures in the radial direction (the ME alone can be up to 0.5~au in radial size at 1~au for very large or expanding events). So, even if CME2 shock arrives 16 hours after CME1 shock, this does not imply the absence of interaction between the two structures. The trailing part of ME1 is, in fact interacting with CME2 in this case.}}

%Magnetic field properties in the sheath - By and Bz features due to interaction
%We first investigate the rotation of the magnetic field in the z-direction ($B_z$) in the sheath ahead of CME2.
Second, we analyse the magnetic field signatures of CME2 in Run4 and highlight the additional features due to the presence of CME1. The first weak drop in $\beta_p$ (proton plasma beta, i.e., $\beta/2$) at $\sim$2014-09-12~06:00~UT is modelled in Run4, while it was missing in Run2, hence reaffirming the short passage of ME1. The drop in $\beta_p$ at $\sim $2014-09-12~19:00~UT in Run4 corresponds to the starting of ME2 and matches the observations well. The period when $\beta_p$ < 1 continues beyond the actual observed end time of ME2 due to the over-expansion of the CMEs in the simulation. The $B_x$ component is not affected much, however, additional structures are observed in the $B_y$ and $B_z$ components in the sheath region ahead of CME2 in Run4 as compared to Run2. A strong drop in $B_y$ component up to $-27\;$nT in the sheath at 2014-09-12 16:00~UT although overestimated, qualitatively corresponds to $B_y = -13\;$nT at $\sim $2014-09-12~18:20~UT in the in situ observations. It is followed by a short-lasting increase to positive $B_y = 9\;$nT at $\sim $~2014-09-12~22:50~UT and a long-lasting negative $B_y$ corresponding to ME2. The following feature, i.e. the transition of the positive $B_y$ to long-lasting negative $B_y$, is sharp in observations. %However, in Run4 the flux rope seems to have expanded more than in reality, leading to a lower magnitude of $B_y$ and a longer passage time of the ejecta. 
The sheath and the magnetic ejecta of CME2 have been reasonably well reproduced in Run4. The sheath has an enhanced positive $B_z = 33\;$nT at $\sim $2014-09-12~14:00~UT ($28\;$nT at 2014-09-12~18:08~UT in situ) followed by a negative $B_z = -5\;$nT ($-17\;$nT at 2014-09-12 20:57~UT in situ) which is clearly missing in Run2. Both simulations capture the prolonged positive $B_z$ in ME2 similarly. The strength of the minimum negative $B_z$ component in the sheath is underestimated at Earth. The virtual spacecraft at $\sigma_{\theta,\phi}$=$\pm$10$\degree$ from Earth registered a minimum $B_z = -11\;$nT around 2014-09-12~19:00~UT, that is closer to the in situ observations. Due to the overestimation of $B_y$ in the sheath, the total magnetic field in the sheath is also overestimated.  

\begin{figure*}
    \centering
    \subfloat[]{\includegraphics[width=0.37\textwidth,trim={0.5cm 1.5cm 1cm 1.5cm},clip=]{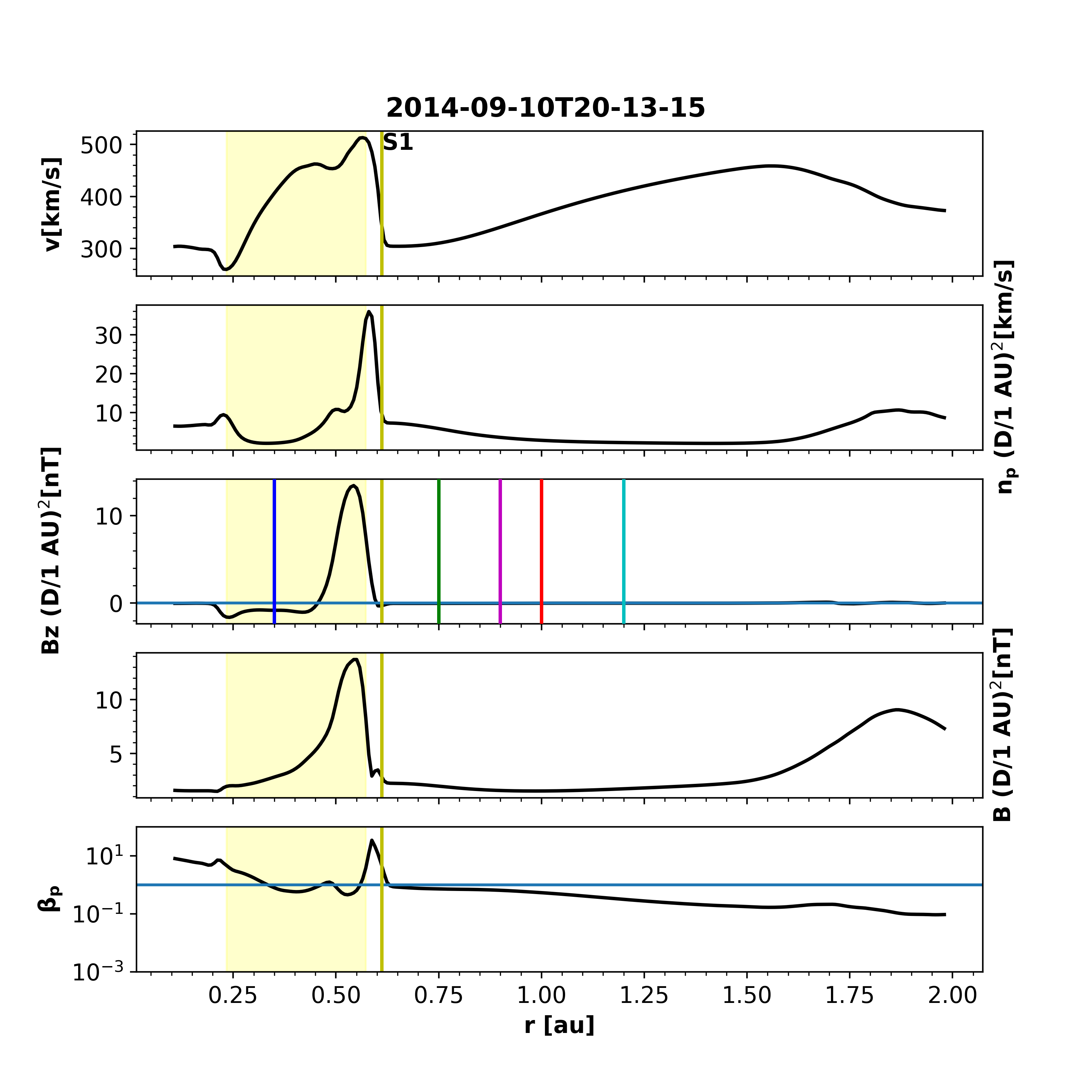}}
    \subfloat[]{\includegraphics[width=0.37\textwidth,trim={0.5cm 1.5cm 1cm 1.5cm},clip=]{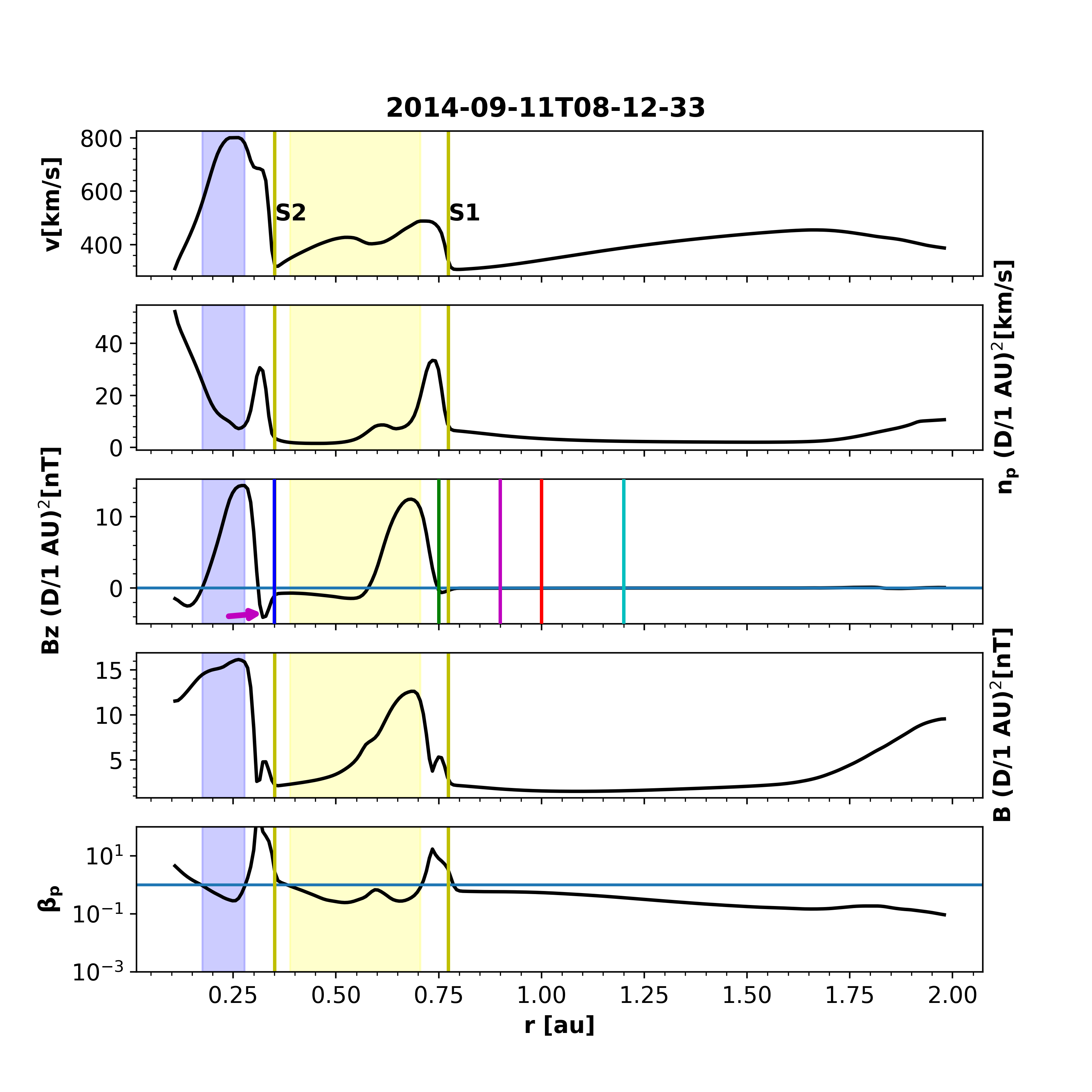}} \\
    \subfloat[]{\includegraphics[width=0.37\textwidth,trim={0.5cm 1.5cm 1cm 1.5cm},clip=]{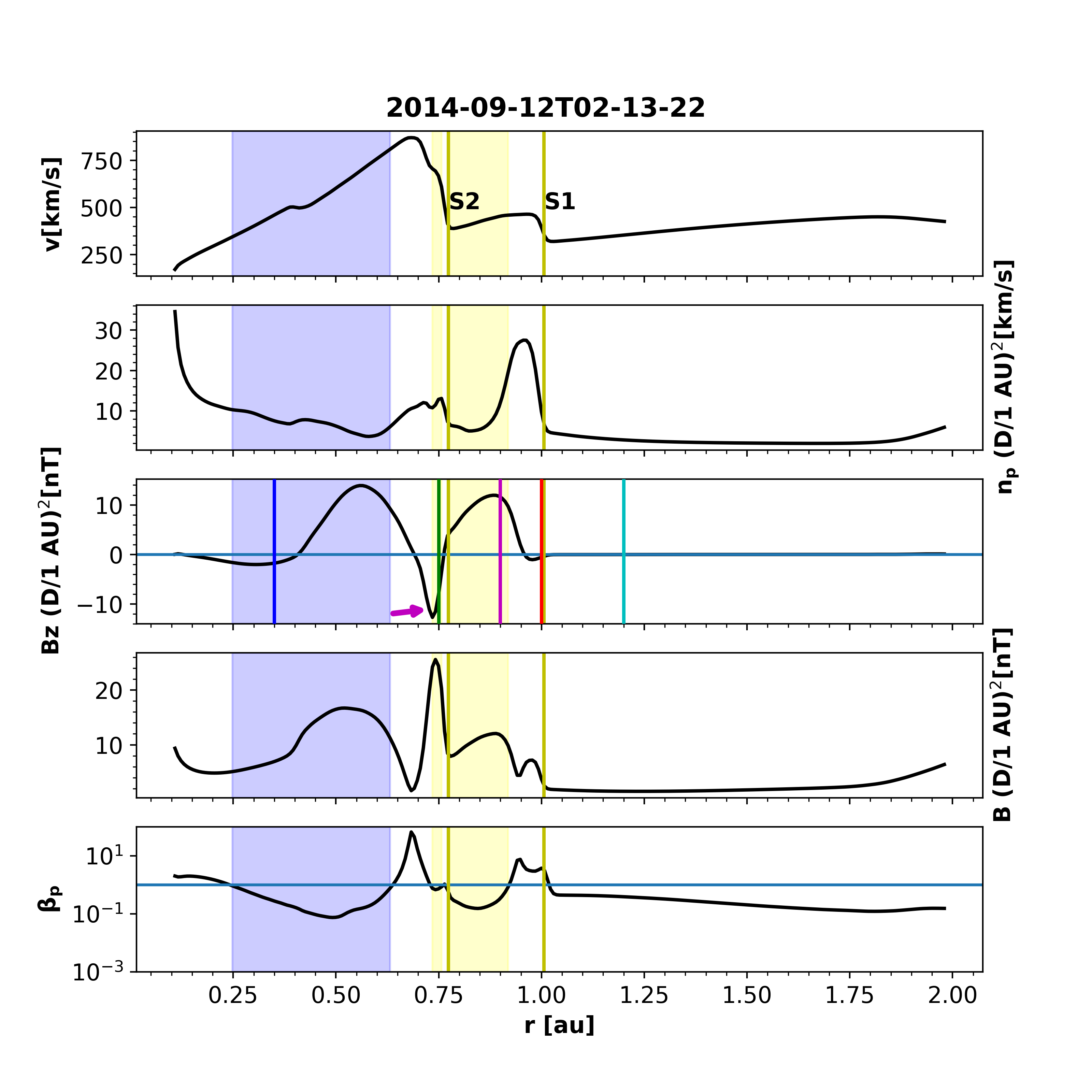}}
    \subfloat[]{\includegraphics[width=0.37\textwidth,trim={0.5cm 1.5cm 1cm 1.5cm},clip=]{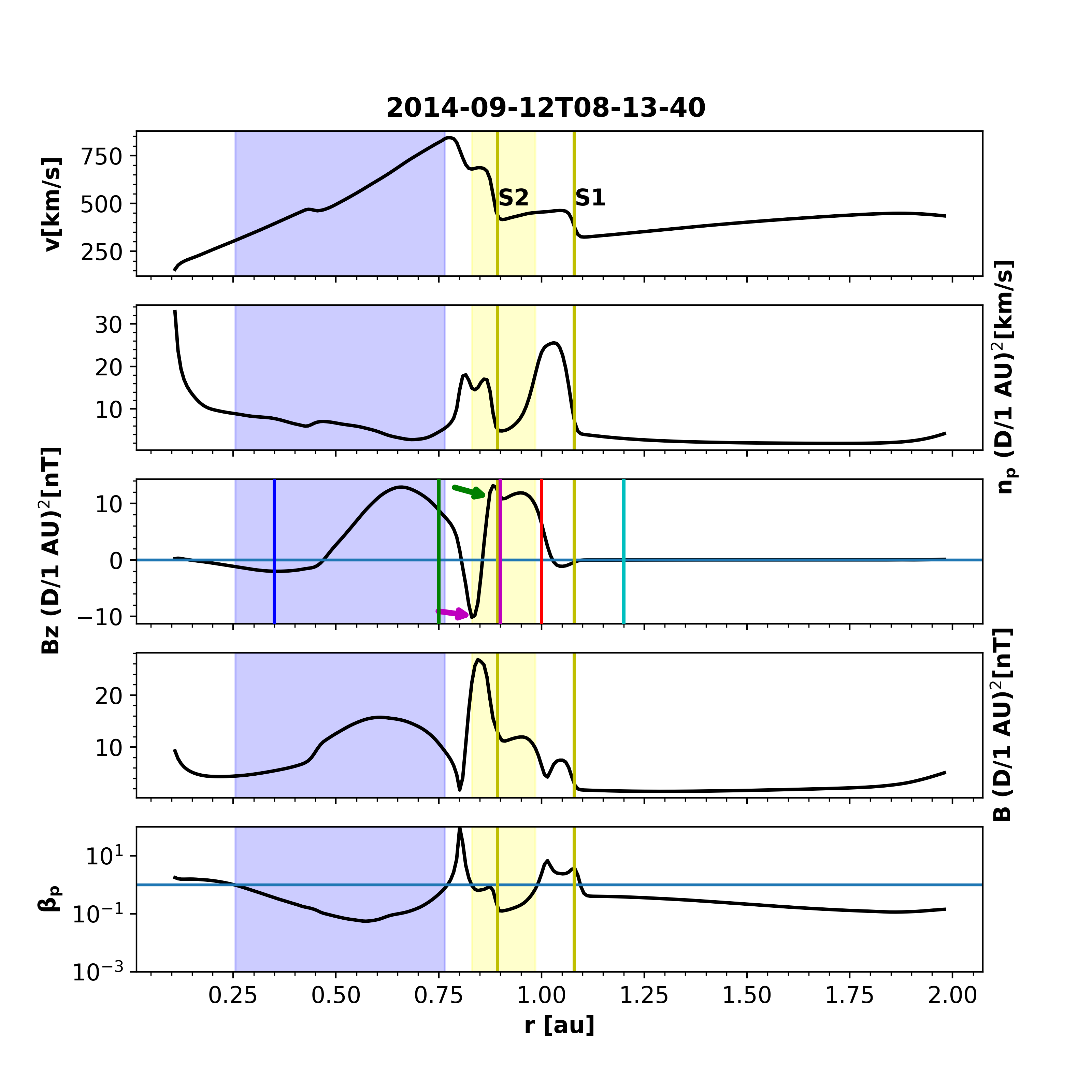}} \\
    \subfloat[]{\includegraphics[width=0.37\textwidth,trim={0.5cm 1.0cm 1cm 1.5cm},clip=]{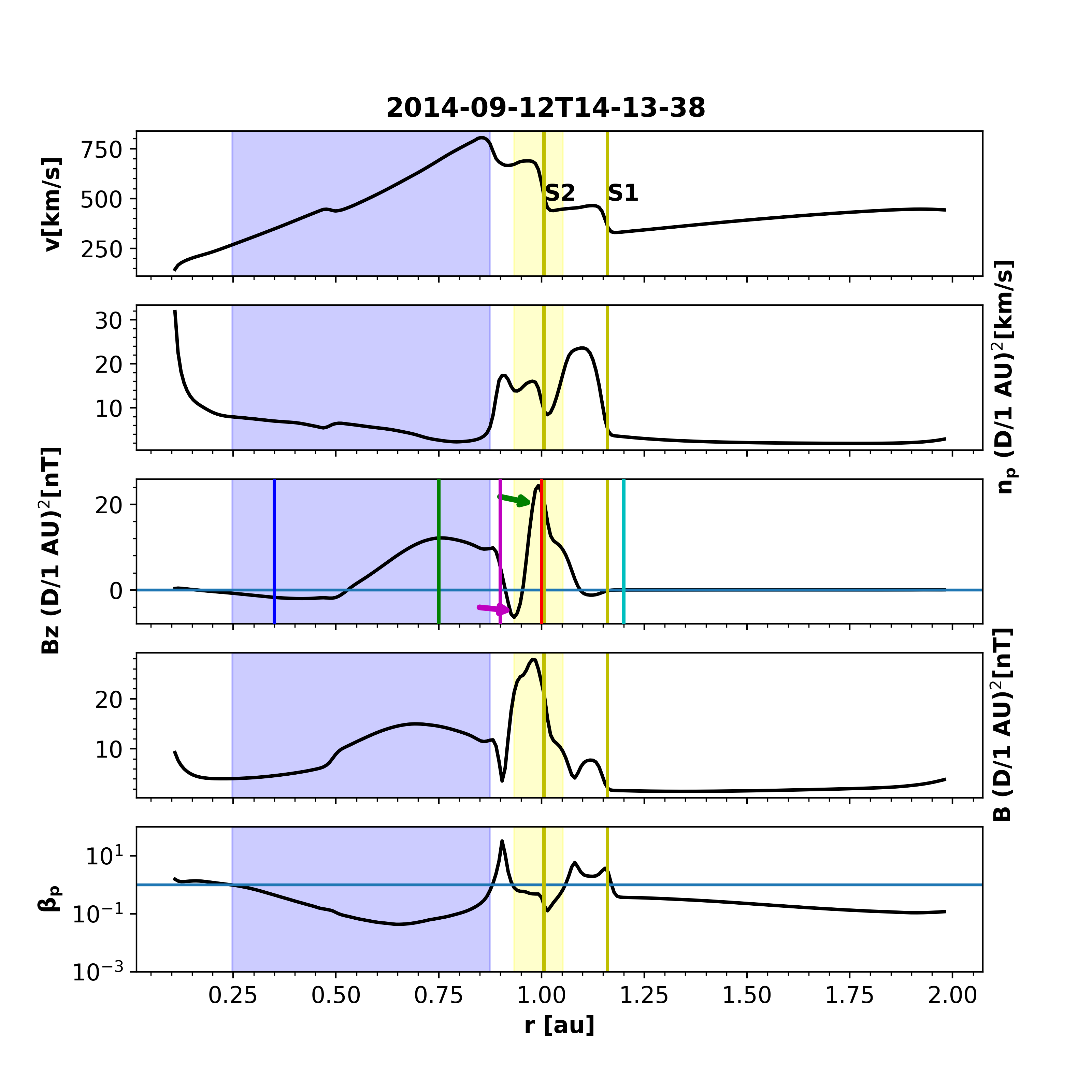}}
    \subfloat[]{\includegraphics[width=0.37\textwidth,trim={0.5cm 1.0cm 1cm 1.5cm},clip=]{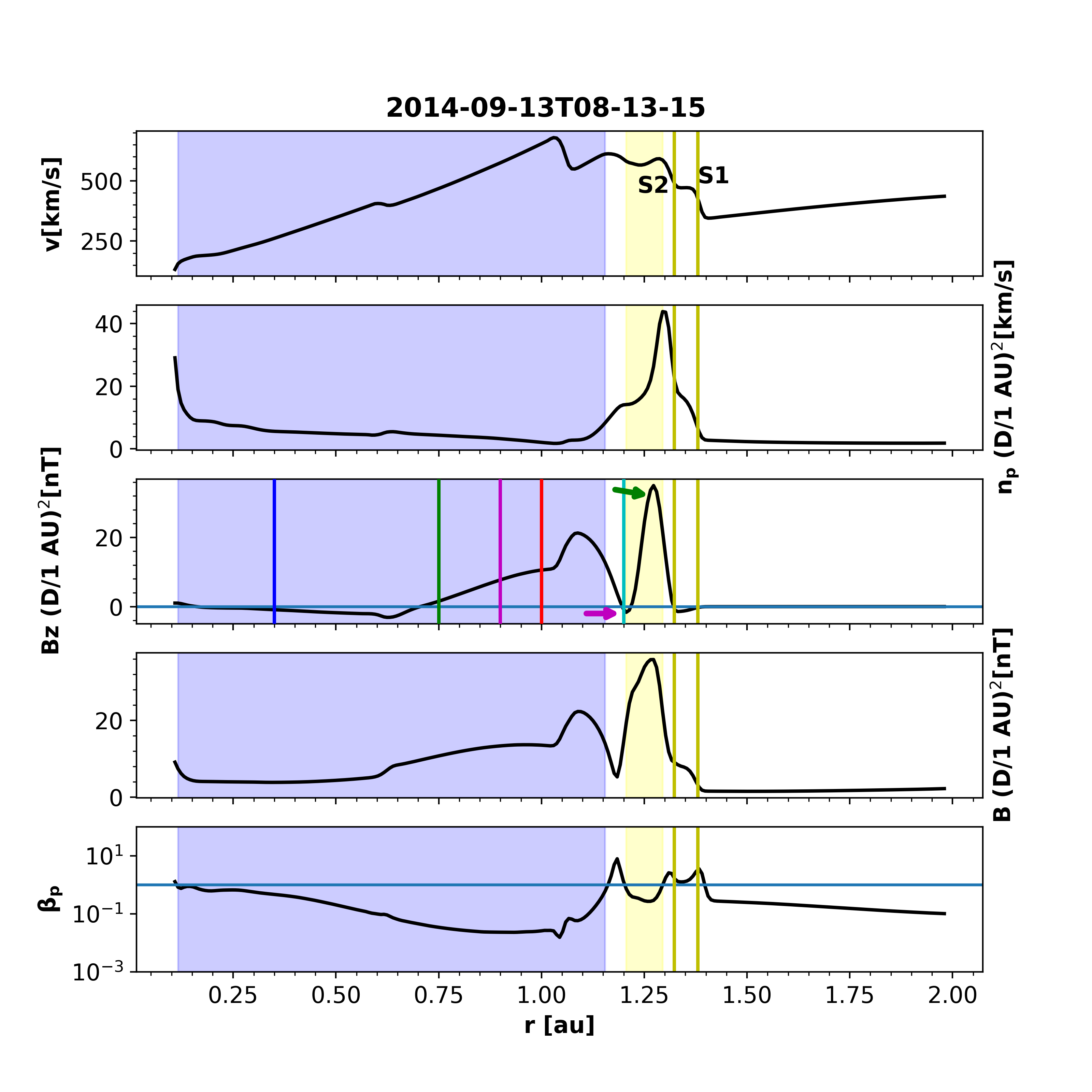}}
    \caption{Radial evolution profile of the CME1 and CME2 along Sun-Earth line at different times of propagation and interaction extracted from Run4. Top to bottom (in each plot): speed ($v$), proton number density ($n_p$), z-component of magnetic field ($B_z$), total magnetic field ($B$), and proton plasma beta ($\beta_p$). The yellow and blue shaded areas depict the extent of magnetic ejecta of CME1 and CME2 respectively extracted using the criterion $\beta_p < 1$ ($\beta_p < 5$ in (a)). All physical quantities except the speed are scaled by (D/1 au)$^2$ where D is the radial distance from the Sun. The CME shocks (S1 and S2) are marked in yellow lines. The colourful vertical lines in the $B_z$ panel correspond to different radial distances to explain various phases of Run4 (detailed in Section~\ref{subsec:cme1+cme2_propagation}): (a) Propagation of CME1 alone in the heliosphere; (b) Formation of a compressed negative $B_z$ ejecta CE1a (shown with magenta arrow); (c) Further compression of CE1a; (d) Development of a compressed positive $B_z$ ejecta, CE1b (shown with green arrow), ahead of CE1a (e) Further compression of CE1b; and (f) Diffusion of CE1a ($\sim0\;$nT) upon reaching $1.2\;$au while CE1b undergoes further enhanced.}
    \label{fig:radial_profiles}
\end{figure*}

\begin{figure*}
    \centering
    %{\includegraphics[width=0.9\textwidth,trim={1cm 5cm 0 0.5cm},clip=]{Figures/bz_eq_mer.pdf}}
    %\caption{Evolution of $B_z$ in the heliosphere as simulated in Run4. $B_{clt}$ is the co-latitudinal component in the spherical coordinate system and is equivalent to -$B_z$ on the ecliptic plane. The red and blue spectra of the colour bar correspond to positive and negative $B_z$, respectively. (a) CME1 in the pre-interaction phase; (b) CME2 evolving behind CME1 in the early stage of interaction; (c) CME2 compressing the trailing negative $B_z$ ejecta of CME1 to create CE1a; and (d) CE1a compressing the leading positive $B_z$ ejecta of CME1 to create CE1b in the interaction phase of the event.}
    {\includegraphics[width=0.9\textwidth,trim={0.5cm 3cm 0 0.5cm},clip=]{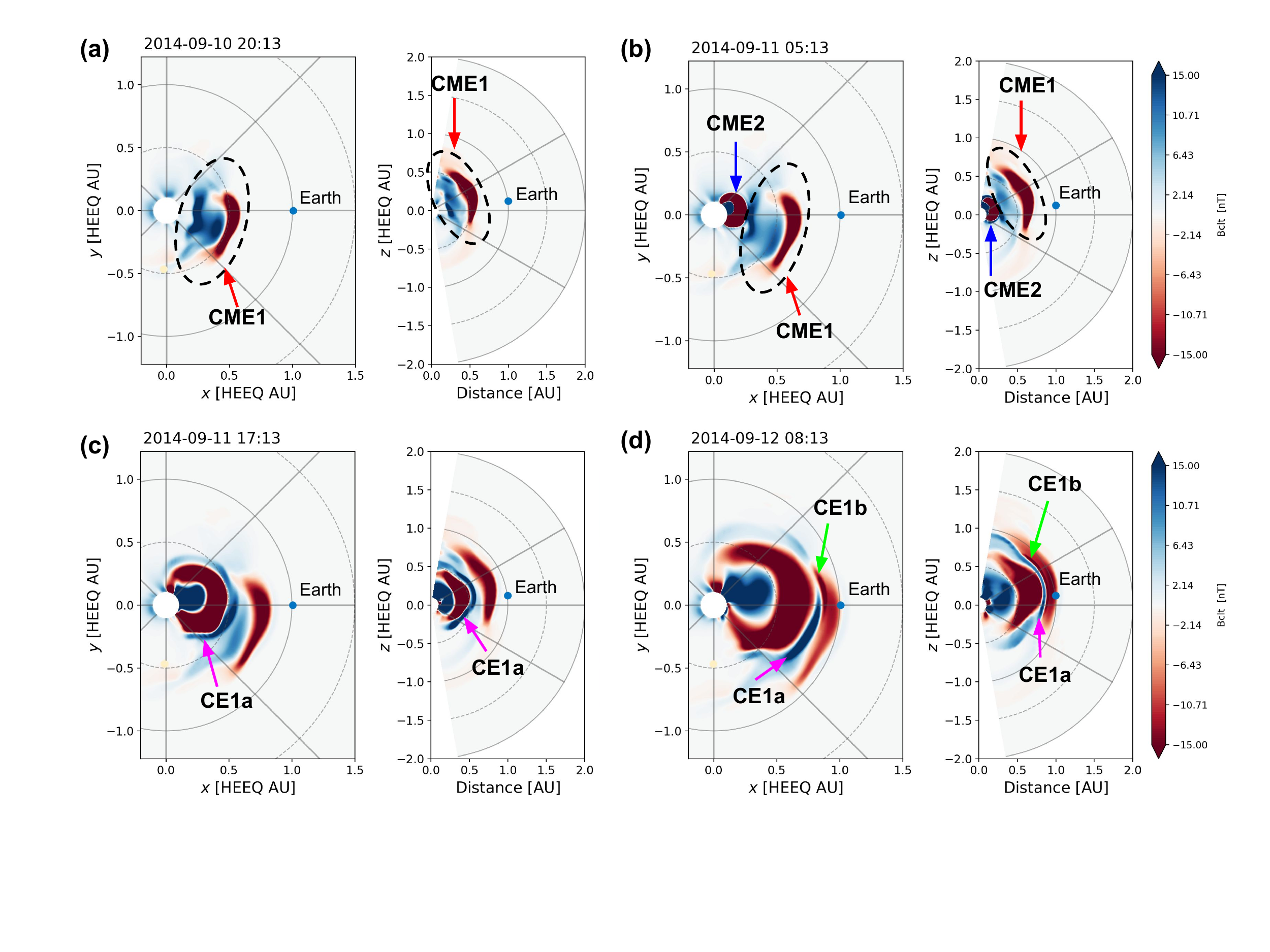}}
    \caption{Evolution of $B_z$ in the heliosphere as simulated in Run4. The co-latitudinal component in the spherical coordinate system is $B_{clt}$, which is equivalent to $-B_z$ on the ecliptic plane. The red and blue spectra of the colour bar correspond to positive and negative $B_z$, respectively. Each sub-figure shows the view of the equatorial (X-Y) and the meridional (X-Z) planes at a particular time mentioned at its top. (a) CME1 in the pre-interaction phase is identified schematically with a dashed ellipse; (b) CME2 is shown evolving behind CME1 in the early stage of interaction; (c) CME2 compressing the interval of the trailing negative $B_z$ component of CME1 to create CE1a (darker blue region) in the sheath ahead of itself; and (d) CE1a is shown to be further compressing the interval of the leading positive $B_z$ component of CME1 to create CE1b (darker red region) during the interaction phase of the event. The animation of this figure can be found in the online version of the paper.
    }
    \label{fig:bz_eq_mer}
\end{figure*}

We have also provided the radial evolution plots for Run4 in Fig.~\ref{fig:radial_profiles}, to understand the evolution of the magnetic ejecta (especially $B_z$) along the Sun-Earth line at different times during the propagation of both the CMEs. $B_{clt}$ (the co-latitudinal component in the spherical coordinate system) is plotted in the equatorial and the meridional plane for Run4 in Fig.~\ref{fig:bz_eq_mer} for a 2D view of the process. $B_{clt}$ is equivalent to $-B_z$ on the equatorial plane. The red and blue spectra of the colour bar correspond to positive and negative $B_z$ respectively. While referring to Fig.~\ref{fig:bz_eq_mer}, we will provide the description in terms $B_z$ instead of $B_{clt}$ to discuss the phenomena. We discuss the results for two phases of the CMEs' propagation process: pre-interaction and interaction. 

\begin{comment}
\begin{figure}
    \centering
    \includegraphics[width=0.5\textwidth]{Figures/CME1+CME2_CME2_NWS.pdf}%newGCS_wind7_fri+45_p+1_lat24_lon15_0p5e14_v500_tw1p5_newGCS_wind7_s-135_v450_lon-14_fri+45_p+1_lat24_lon15_0p5e14_v500_tw1p5.png}%{Figures/20140910_newGCS_wind2_fri+65_p+1_Earth_WIND.png}
    \caption{Results of Run2 (blue) and Run4 (red) are compared. CME1 and CME2 parameters in Run4 correspond to Run3 and Run2 respectively.}
    \label{fig:sim_cme1_cme2_non_marked}
\end{figure}    
\end{comment}

\textit{Pre-interaction}: Run4 gives similar results to Run3 in the pre-interaction phase until 2014-09-12~12:33~UT. In Fig.~\ref{fig:bz_eq_mer}(a), the leading part of the magnetic ejecta associated with CME1 can be observed to propagate with a positive $B_z$ component followed by a negative $B_z$ component, and the same signatures are reproduced in situ when ME1 arrives at Earth (Fig.~\ref{fig:sim_cme1_cme2}).  

\textit{Interaction}: In Fig.~\ref{fig:radial_profiles}, the vertical yellow lines corresponding to the shock of the CMEs (S1 and S2), and the other coloured vertical lines in the $B_z$ panel marking different locations (radial distances) from the Sun are used for the purpose of description in this section. {{The yellow and blue shaded areas mark the extent of magnetic ejecta of CME1 and CME2 respectively extracted using the criterion $\beta_p < 1$ in all figures except for Figure~\ref{fig:radial_profiles}(a). The criterion of low $n_p$, in combination with $\beta_p < 5$ is used to identify the CME1 extent below $0.5\;$au in Figure~\ref{fig:radial_profiles}(a).}} It shows the phase where CME1 has propagated alone in the heliosphere with a dominant positive $B_z$ component ($>10\;$nT), followed by a weak minimum negative $B_z$ component ($\sim -2\;$nT), and has reached $0.65\;$au at 2014-09-10~20:13~UT. The extent of CME1 in the equatorial and meridional planes of the heliospheric domain can be observed in Fig.~\ref{fig:bz_eq_mer}(a). In the next phase in Fig.~\ref{fig:radial_profiles}(b), CME2 has entered the heliospheric domain and its shock has reached $0.35\;$au (marked by yellow lines). An enhancement of the negative $B_z$ component interval of the ejecta (hereafter, compressed ejecta 1, CE1a) to $\sim -4\;$nT is observed in the CME2 sheath at 2014-09-11~08:13~UT just before $0.35\;$au (vertical blue line). This enhancement is present in the sheath region as the corresponding $\beta_p > 1$ and can be interpreted as the interval of trailing negative $B_z$ component of CME1 being compressed by CME2. {{CME1 appears weaker than CME2 as it is the flank of the CME1 that hits Earth and the magnetic field strength of a flux rope becomes weaker away from its axis. In our simulations, we observe that CME1 expands rapidly in the heliosphere and hence, has a larger trailing part extending up to $\sim0.2$~au while the leading part has reached $\sim0.6$~au (Fig.~\ref{fig:bz_eq_mer}(b)). Hence, CME2 (faster than CME1) could catch up with CME1 and start compressing it below 0.5~au.}} Fig.~\ref{fig:bz_eq_mer}(c) shows the prominent development of CE1a at 2014-09-11~17:13~UT. Figure~\ref{fig:radial_profiles}(c) shows CE1a (marked with magenta arrow in $B_z$ panel) in the CME2 sheath being further compressed to $B_z < -10\;$nT as it reaches $0.75\;$au at 2014-09-12~02:13~UT while CME1 shock reaches $\sim 1\;$au (vertical red line). A strong positive $B_z > 10\;$nT in the region of $\beta < 1$ up to $0.7\;$au corresponds to ME2 which seems to be pushing CE1a further.
The next phase depicted in Fig.~\ref{fig:radial_profiles}(d) is the development of an enhancement in the interval of the leading positive $B_z$ component of CME1 ahead of CE1a (hereafter, compressed ejecta 2, CE1b, marked with green arrow in $B_z$ panel) at $0.9\;$au (vertical magenta line) at 2014-09-12~08:13~UT. Through Fig.~\ref{fig:bz_eq_mer}(c) and (d), it can be inferred how CE1a compresses the interval of the leading positive $B_z$ component of ME1 to create CE1b. It must be noted that these features are very thin and localised.
Fig.~\ref{fig:radial_profiles}(e) shows the further enhancement of CE1b at $1\;$au with a $B_z > 20\;$nT which is more than the maximum $B_z \sim 10\;$nT inside ME1. CE1a starts weakening and has a lesser magnitude at 2014-09-12~14:13~UT as compared to the in situ observations. However, when CE1a was at $0.9\;$au the minimum negative $B_z$ matched the $1\;$au observations better. In addition, it is evident from Fig.~\ref{fig:bz_eq_mer}(d) that the most enhanced part of CE1a (\ie \ minimum $B_z$ component) is to the east of the Sun-Earth line and to the north of the ecliptic. The enhanced features are quite small compared to the prominent magnetic ejecta in the event and are localised enough to be missed easily while reading out the 3D data at a single point in the simulation.
Fig.~\ref{fig:radial_profiles}(f) shows the phase where CE1a has a very weak magnetic field strength upon reaching $1.2\;$au (vertical cyan line) at 2014-09-13~08:13~UT while CE1b is further compressed. CE1a seems to have been compressed between ME2 and CE1b. {{The propagation of the CME2 shock (the trailing vertical yellow line) towards the CME1 shock (the leading vertical yellow line) can be observed through Fig.~\ref{fig:radial_profiles}(c-f). The CME2 shock moves across the $\beta < 1$ region associated with CME1, along the Sun-Earth line which can be inferred as a signature of interaction. Moreover, the $\beta_p < 1$ part associated with CME1 gets narrower in time, which points to the subsequent increase in compression by CME2.}}

\begin{comment}
\begin{figure}
    \centering
    \includegraphics[width=0.5\textwidth]{Figures/20140910_newGCS_wind2_fri+65_p+1_20140910_newGCS_wind2_spr-135_fri+65_p+1_Earth_WIND.png}
    \caption{Comparing red run with CME1 and CME2, and blue run with only CME2.}
    \label{fig:spr-135_fri+65+1}
\end{figure}
\end{comment}

\begin{comment}
\begin{figure}
    \centering
    \includegraphics[width=0.5\textwidth]{Figures/R09_20140910_newGCS_wind7_s-135_v450_lon-14_fri+45_p+1_lat24_lon15_0p5e14_v500_tw1p5.png}%{Figures/20140910_newGCS_wind2_spr-135_fri+65_p+1_20140910_newGCS_wind2_spr-135_fri+65_p+1_Earth_-8h_WIND.png}
    \caption{Time series of Run4 taken at r=0.9~au in solid blue line. The shaded time series correspond to $\sigma_{\theta,\phi}$}
    \label{fig:Run4_R09}
\end{figure}
\end{comment}

%$B_{clt}$ figures are from /media/u0141347/T7/EUHFORIA/20140910_newGCS_wind2_s-135_v450_lon-14_fri+45_p+1_lat24_lon15_0p8e14_v525/heliosphere

%trim={1cm 1.25cm 1cm 1.5cm} for the no_shade... pngs

%\subsection{CMEs in a uniform wind}
%The next set of simulations are performed in uniform solar wind background in order to investigate the role of interaction of the CMEs with the high speed stream in the solar wind.

%%%---------------------%%%%%%%%%%%%%%%%%%%%%-------------------%%%
\subsection{Parameters affecting the sheath formation}
\label{subsec:discussion}
%\textit{(1) Discuss the factors affecting the modelling of the distinct features due to interaction. Focus on relative speed, magnetic field orientation and strength.}
%The compression dynamics can be explained similarly using the $B_y$ component of the magnetic field as well.

%Through our numerical experiments, we observed that the factors such as the relative speed, relative direction and time of launch, and most importantly the magnetic field orientation (also studied by previous works e.g., \citet{Lugaz2013,Kilpua2019}) played a crucial role in obtaining the appropriate interaction region between the two CMEs. %The relative speed between the two CMEs is crucial in determining the distinct magnetic field structures in the sheath ahead of CME2 (positive $B_z$ followed by negative $B_z$).

The width and the duration of the compressed features in the sheath of CME2 (the CME that propagates behind CME1 and compresses it while catching up with it) depend on the relative speed between the two CMEs and on the physical properties of both CMEs, such as their shape and size \citep{Russell2002}. Additional simulations were performed (whose detailed results are not shown in this paper) to analyse the effect of these parameters on the formation of the features in the sheath ahead of CME2. The duration of a particular feature, for example, CE1b, depends on the compression induced by CE1a on the $B_z$ profile (with intervals of negative and positive values) of CME1 during their propagation through the interplanetary space. 
%If CME2 is faster (i.e., higher relative speeds), it can catch up with CME1 earlier and can compress the negative $B_z$ part of CME1 more, generating a thinner and diffused CE1b by the time they arrive at $1\;$au. We also find in our simulations that, for a slower CME2 (i.e., lower relative speeds), CE1a catches up with the positive $B_z$ part of CME1 to form CE1b at a later time. In this case, the compression takes place beyond the Earth's orbit and the simulation fails to capture the enhanced positive $B_z$ signature in the sheath before the negative $B_z$ signature. 
If CME1 is slower (i.e., in the case of a smaller relative speed), CME2 can catch up with it earlier and can compress the interval with the negative $B_z$ component of CME1 more, generating a thinner and diffused CE1b by the time they arrive at $1\;$au. We also find in our simulations that, for a faster CME1 (i.e., in the case of a higher relative speed), CE1a catches up with the interval with the positive $B_z$ component of CME1 to form CE1b at a later time. In this case, the compression takes place beyond the Earth's orbit, and the simulation fails to capture the enhanced positive $B_z$ signature in the sheath before the negative $B_z$ signature.
Given the short spatio-temporal nature of these distinct features, these dynamics can be easily missed if virtual spacecraft are not taken into consideration. The minimum $B_z$ strength of CE1a at $1\;$au is indeed captured by the spacecraft at 10$\degree$ longitudinal offset from Earth rather than the solid red line at Earth in Run4 (Fig.~\ref{fig:sim_cme1_cme2}). The magnetic field strength of the compressed ejecta depends on the flux contained in CME1. Moreover, if the magnetic field configuration of CME1 was not modelled correctly, then the $B_y$ and $B_z$ features in the sheath would have had different signs and may have resulted in completely different features in the sheath. The shape and size of CME1 also play an important role in the morphology of the compressed ejecta. %As we are modelling the flank encounter by CME1 with the spheromak model which has the drawback of its compact spherical shape, we had to shift it longitudinally (by $\sim 4\degree$) towards the Sun-Earth line to capture the interaction effects. 
The flux and twist of CME2 not only influence its magnetic field strength obtained at 1~au but also its ability to compress the ejecta ahead. This event clearly depicts how the sheath region carries the history of the minute interactions between different magnetic ejecta, which might be challenging to predict based on remote-sensing observations. Although 3D MHD simulations help in modelling and understanding such interactions better, it can still be challenging to reproduce the observed features exactly, owing to the sensitivity of the simulations to the uncertainties in the initial parameters as discussed above.

%%%---------------------%%%%%%%%%%%%%%%%%%%%%-------------------%%%
\section{{Summary and conclusions}}
\label{sec:conclusion}
%%%---------------------%%%%%%%%%%%%%%%%%%%%%-------------------%%%
%Story: Why Investigate the puzzling misprediction?
In this study, we presented the evolution of two successive CMEs that erupted from the active region AR 12158 on September 8, 2014, and September 10, 2014, respectively. %from AR 12158. %a side hit on Earth and provided preconditioning in the heliosphere for the second CME’s propagation. 
%A detailed motivation for studying this event was provided in Section~\ref{sec:event_overview}.
The motivation of this work was to investigate the misinterpretation of certain observational aspects related to the CMEs and to reproduce their in situ signatures at $1\;$au using MHD simulations. % by correcting the initial parameters.
%(see for  details Section~\ref{sec:event_overview}).}
The first CME was not predicted to hit Earth and was not even recorded in the ICME catalogues. The second CME was predicted to be geoeffective based on the remote observations of the CME chirality and magnetic axis orientation during the eruption. However, unexpectedly, upon arrival at Earth its magnetic ejecta was dominated by a positive $B_z$ component. Nonetheless, a short period of negative $B_z$ component developed in the sheath of CME2 during its propagation in the heliosphere. This resulted in a geomagnetic storm with a Dst index of $\sim -88\;$nT at Earth, which is a moderate value and not an extreme value as was originally predicted. Hence, the geoeffectiveness of the various sub-structures involved in this event was gravely mispredicted. The study of this event is based on the observational investigation of the CMEs, and the 3D MHD modelling of their evolution using EUHFORIA. %Observational work: Track orientation from Sun to Earth
We performed an in-depth analysis of the two CMEs using remote sensing and in situ observations, and constrained the geometrical and magnetic field parameters successively, close to $1\;$R$_\odot$, close to 0.1~au and at 1~au. A discrepancy was observed in the axial magnetic field orientation of CME2 between $1\;$R$_\odot$ and $0.1\;$au. {{The numerical experiments for involving CME1 and CME2 individually (Run1, Run2, Run3) and the global EUHFORIA simulation including CME1 and CME2 (Run4) have the following rationales:}} %(as presented in Section~\ref{sec:obs_analysis_remote} and \ref{sec:obs_analysis_insitu}).

\begin{itemize} 
    \item Run1: We first performed a simulation including only CME2 using the FRi3D model in EUHFORIA, with the magnetic field orientation obtained close to $1\;$R$_\odot$. However, the results did not match the in situ magnetic field observations at $1\;$au. 
    \item Run2: In order to determine the cause of this discrepancy, we performed another simulation with the initial magnetic field orientation consistent with in situ observations at $1\;$au using the FRi3D model in EUHFORIA. The inference of the new magnetic field orientation of CME2 is based on the 3D reconstruction (close to 0.1~au) and the assumption of anti-clockwise rotation of left-handed CME2 in the low corona (based on the relationship between the chirality and the rotation direction from \citet{Green2007,Lynch2009}). Run2 could reproduce the prolonged positive $B_z$ component seen in the in situ observations and matched the observations much better than Run1. However, Run2 requires a significant $\sim 180\degree-270\degree$ rotation of CME2 in the low corona which remains to be explained, as no rotation was observed during the heliospheric propagation in the simulations. 
    \item Run3: After obtaining the best simulation of CME2, we performed a simulation including just CME1 with the spheromak model. The magnetic field orientation of CME1 close to $1\;$R$_\odot$ was consistent with the orientation obtained close to 0.1~au and was used to obtain the best simulation of CME1 to reproduce the in situ signatures at $1\;$au.
    \item Run4: Finally, we introduced CME1 and CME2 %(as in Run3) before CME2 (as in Run2) in the simulation domain 
    {{using the boundary conditions found in Run3 and Run2 to create the final global simulation (Run4).}} The kinematics and the magnetic field components of the CMEs were successfully modelled at Earth with the magnetised flux rope CME models in EUHFORIA. The interaction between CME1 and CME2 was found to produce the short interval of negative $B_z$ component in the sheath ahead of CME2. With the 3D MHD EUHFORIA simulation, it was possible to understand the different phases of CME1-CME2 interaction forming coherent enhanced sub-structures (CE1a and CE1b) in the sheath region.% as discussed in Section~\ref{subsec:cme1+cme2_propagation}. 
\end{itemize}

In a nutshell, we investigated the reasons for the space weather misprediction of this event. We found it to be two-fold: first, the consideration of a low coronal rotation of CME2 was missed which led to predicting a different magnetic field topology heading towards Earth; second, the presence of CME1 was overlooked and hence the geoeffective feature formed by its interaction with CME2 was not predicted. {{EUHFORIA in its present version allows us to propagate the CMEs in the heliosphere. We have shown that a substantial rotation of CME does not take place in the heliosphere. Therefore, we suggest that the rotation of CME2 could have occurred in the corona.}} With this study, we also highlight the importance of observations in correctly constraining simulations for obtaining accurate space weather forecasting.\\

Previous studies suggest that a significant amount of CME rotation, deflection, and deformation occurs in the low corona, followed by a self-similar propagation further away from the Sun in most CME events \citep{demoulin2009,Isavnin2014,balmaceda2020}. \citet{Kliem2012} have highlighted the possibility of extensive low coronal rotation up to even more than 100$\degree$ by the combination of twist and shear-driven rotation, the latter being dominant in the lower corona. It is not possible to verify the hypothesis of a substantial rotation ($\sim 180\degree-270\degree$) of CME2 in the low corona under the scope of this work as EUHFORIA does not account for the modelling of the initial CME evolution below 0.1~au. In addition, the lack of magnetic field observations in the corona restricts the observational verification of this hypothesis. This speculation, although a relatively strong assumption, provides an idea about the consistency of the CME2 orientation observed in the upper corona and at 1~au. Hence, the knowledge of the CME magnetic field is crucial for deriving the correct orientation of the emerging flux rope in the low corona, in order to propagate it further in the heliospheric models for space weather forecasting purposes. %The main limitation that we face is the lack of magnetic field observations of the CMEs in the corona. 
Although the white light coronagraph images help to reconstruct the CMEs in the middle and upper corona, they are not sufficient to derive the magnetic field configuration. This leaves us to rely on the source region proxies for guessing the magnetic field configuration for prediction purposes, being agnostic to the dominant low coronal dynamics. Although the closest approach of Parker Solar Probe is in the upper corona, its trajectory does not act as a constant in situ monitoring point in the corona. The other forecasting limitation arises from the evolution of CME structures during their propagation, and their interaction with the solar wind and other CMEs. The interaction of CMEs may lead to severe geoeffective events, as demonstrated by \citet{Shen2018},\citet{Scolini2020} and \citet{Koehn2022}. The lack of multiple in situ crossings through CME2 makes it challenging to predict its global behaviour through reconstruction using data from just a single point. Furthermore, for CME crossings with a high impact factor, single-point reconstruction techniques introduce greater uncertainty in the estimation of the flux rope orientation \citep{Riley2004}. The serendipitous alignment of spacecraft like PSP, Solar Orbiter, and Bepi Colombo, although helpful in obtaining information about the early phase of the CME, is also not feasible for constant monitoring.  Without the knowledge of the global behaviour of the individual CMEs, it is even more challenging and non-trivial to predict the strength and configuration of the magnetic ejecta formed during CME-CME interaction. Hence, we advocate for a stronger observational infrastructure for the study of CME-CME interaction events from the perspective of space weather forecasting, in addition to MHD simulations.

%%%---------------------%%%%%%%%%%%%%%%%%%%%%-------------------%%%
\section*{Acknowledgements}
%%%---------------------%%%%%%%%%%%%%%%%%%%%%-------------------%%%
We thank the anonymous referee for their comments and suggestions that led to improvements in the manuscript. This project (EUHFORIA 2.0) has received funding from the European Union’s Horizon 2020 research and innovation programme under grant agreement No 870405. SP acknowledges support from the projects
C14/19/089  (C1 project Internal Funds KU Leuven), G.0D07.19N  (FWO-Vlaanderen), SIDC Data Exploitation (ESA Prodex-12), and Belspo project B2/191/P1/SWiM. The simulations were carried out at the VSC – Flemish Supercomputer Centre, funded by the Hercules Foundation and the Flemish Government – Department EWI. We are grateful to Dr. Nariaki Nitta and Dr. Tibor Torok for the valuable discussions that improved our understanding of the eruptions. We thank Dr Jasmina Magdaleni\'c for the suggestions that facilitated the better 3D reconstruction of the halo CMEs. We also appreciate the availability of open-source data and catalogues used in this work:
\begin{itemize}
    \item CDAW LASCO catalogue: \url{https://cdaw.gsfc.nasa.gov/CME_list/}
    \item IPShock catalogue: \url{http://ipshocks.fi/}
    \item Wind ICME catalogue: \url{https://wind.nasa.gov/ICME_catalog/ICME_catalog_viewer.php} 
    \item Richardson and Cane ICME catalogue: \url{https://izw1.caltech.edu/ACE/ASC/DATA/level3/icmetable2.htm}
    \item AIA images: SDO database \url{http://jsoc.stanford.edu/AIA/AIA_gallery.html}

\end{itemize}

\appendix
\section{FRi3D model flux calculation}
\label{append:FRi3D_flux}
The toroidal flux ($\phi_t$) as a function of the poloidal flux ($\phi_p$) for a flux rope with Lundquist magnetic field configuration is given by \citep{Gopalswamy2018}:
\begin{equation}
    \phi_t = \phi_p \frac{2\pi R_0}{L} J_1(x_{01})
\end{equation}
where $R_0$ and $L$=2.6 $R_{tip}$ are the radius and length of flux rope respectively. $R_{tip}$ is the leading edge of the flux rope. $J_1$ is the first order Bessel function and $x_{01}$ is the first zero of zeroth order Bessel function, $J_0$.
We modify above formula for FRi3D geometry by replacing $R_0$ with the poloidal height of FRi3D ($R_p$) and $L$ with the FRi3D axis length given by:
\begin{equation}
    L = \int_{-\phi_{hw}}^{\phi_{hw}} \bigg[r(\phi)^2 + (\frac{dr(\phi)}{d\phi})^2\bigg]^{\frac{1}{2}} \,d\phi
\end{equation}
where $r(\phi) = R_t cos^n(a\phi)$ is the cross-section at a given $\phi$ and $a=(\pi/2)/\phi_{hw}$. $R_t$ is the toroidal height and $\phi_{hw}$ is the angular half-width. Further details of FRi3D geometry and the parameters used here can be found in \citet{Maharana2022}.

\section{3D reconstruction of the CMEs}
\begin{figure}
    \centering
    \includegraphics[width=0.5\textwidth,trim={4.5cm 1.5cm 5cm 1.5cm},clip=]{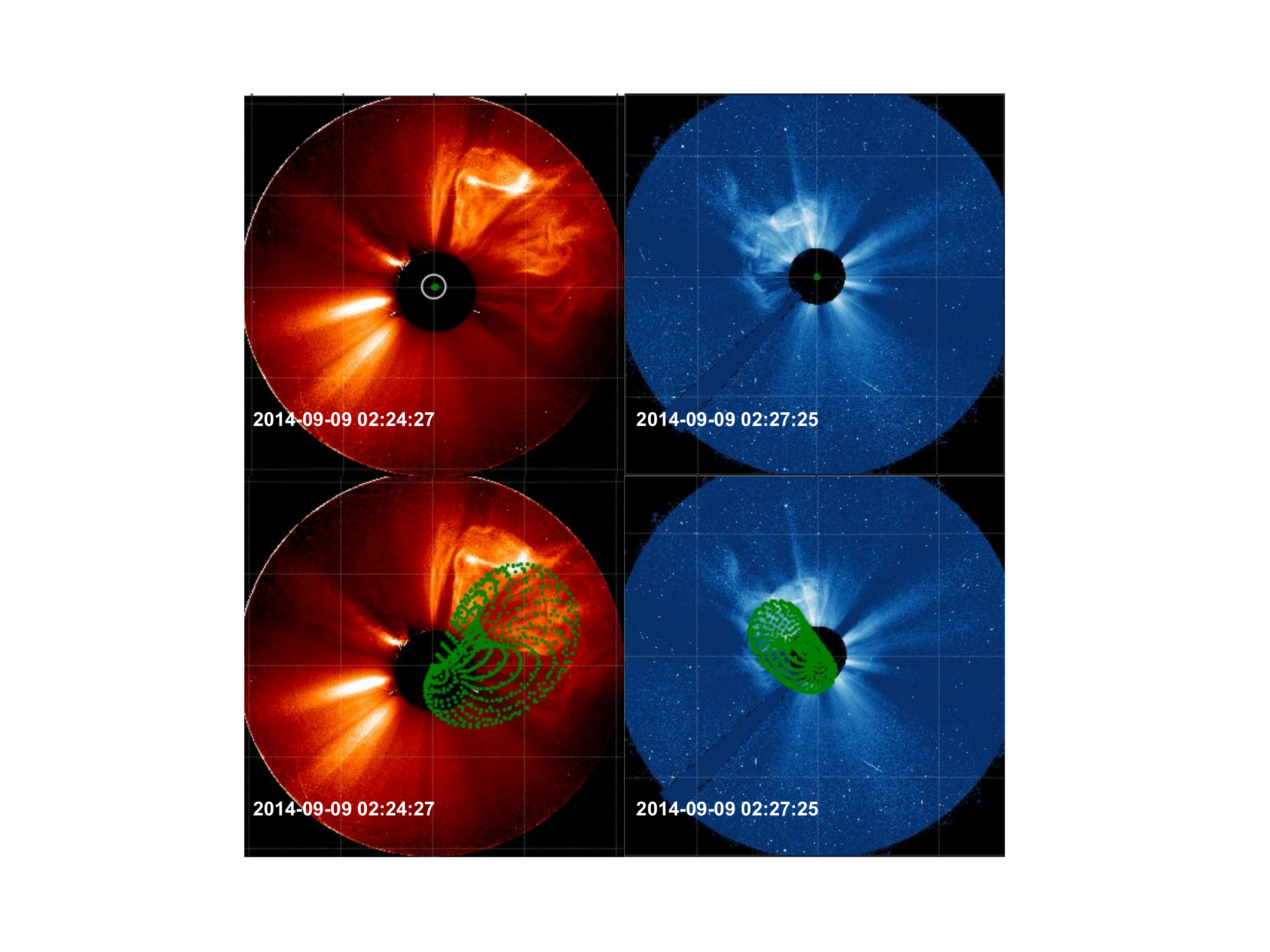} \\%CME1_GCS_20140909_0224.pdf}
    \caption{Coronagraph images of CME1 development on September 9, 2014, as observed by COR-2 (STEREO-B) and C3 (LASCO) (top). The same images overlaid with the 3D reconstruction using the GCS model (bottom)}
   \label{fig:cme1_3d_fit}
\end{figure}

\begin{figure}
    \centering
    \includegraphics[width=0.4\textwidth,trim={0cm 0cm 0cm 0cm},clip=]{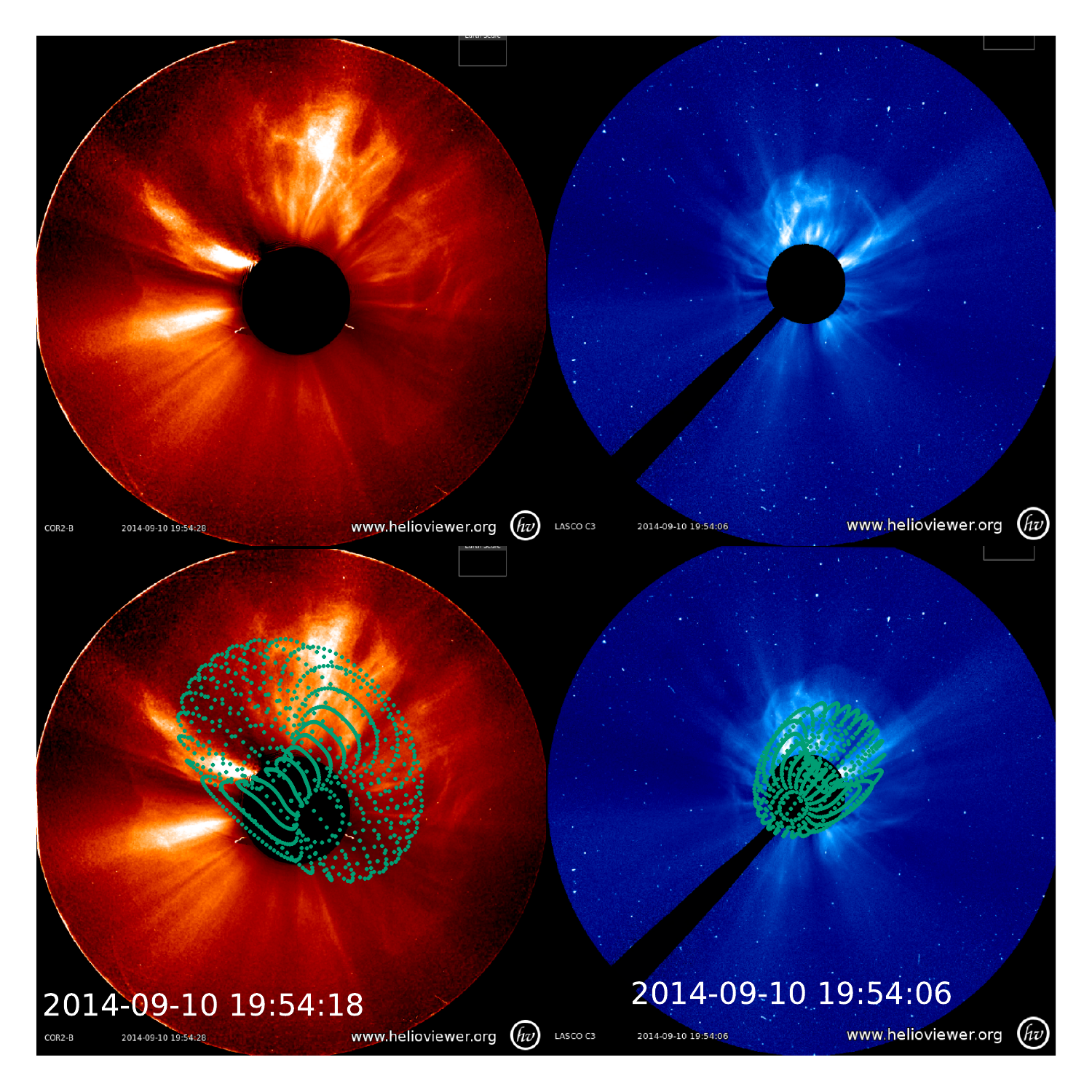}
    \caption{Coronagraph images of CME2 development on September 10, 2014 as observed by COR2 (STEREO-B) and C3 (LASCO) (top). The same images overlaid with the 3D reconstruction using the FRi3D model (bottom).}
   \label{fig:cme2_3d_fit}
\end{figure}

The geometrical and kinematic parameters of CME1 and CME2 are constrained from the 3D reconstruction using the GCS model \citep{Thernisien2011} and the FRi3D model \citep{Isavnin2016ApJ}, respectively. As CME1 is modelled with the spheromak model in EUHFORIA simulations, it is reconstructed with the GCS model (Fig.~\ref{fig:cme1_3d_fit}) as done in \citet{Verbeke2019,Scolini2019}. CME2 is reconstructed with FRi3D model (Fig.~\ref{fig:cme2_3d_fit}) and modelled with the same in EUHFORIA.

\end{document}